\documentclass{JHEP3}
\usepackage{latexsym,bm,amsmath,amssymb,amsfonts, dsfont}
\usepackage{epsfig,graphics,graphicx}
\usepackage{mathrsfs}
\usepackage{slashed}
\usepackage{cite}
\usepackage[latin1]{inputenc}

\newcommand{\dif}{\mathrm{d}}
\newcommand{\mcal}{\mathcal}
\newcommand{\mcN}{\mathcal{N}}
\newcommand{\mcF}{\mathcal{F}}
\renewcommand{\vec}[1]{\boldsymbol{#1}} 
 \newcommand{\abar}{\bar{\alpha}_s}
  
  \newcommand{\CKT}{\tilde{\mathcal{K}}}
  \newcommand{\ktp}{{\vec{k}}^{\prime}}
  \newcommand{\ktpsq}{{k'}^2}        
  \newcommand{\kto}{{\vec{k}}_0}
\newcommand{\kc}{\mathrm{kc}}           
\newcommand{\qt}{\vec{q}}        
\newcommand{\asb}{\overline{\alpha}_s}
\newcommand{\ci}{{\mathrm{c}}}
\DeclareMathOperator{\Li}{Li_2}         
\newcommand{\kt}{\vec{k}}       
\newcommand{\om}{\omega}
\newcommand{\omhalf}{\tfrac{\omega}{2}}

\title{\rm Next-to-leading and resummed BFKL evolution with saturation boundary}
\author{E.~Avsar$^{a}$, A.M.~Sta\'sto$^{a,b,c}$, D. N.~Triantafyllopoulos$^{d}$, D.~Zaslavsky$^{a}$ \\
\!\!$^{a}$104 Davey Lab, Penn State University, University Park, 16802 PA,  USA\\
\!\!$^{b}$RIKEN Center, Brookhaven National Laboratory, Upton, NY 11793, USA \\
\!\!$^{c}$Institute of Nuclear Physics, Polish Academy of Sciences, Cracow, Poland \\
\!\!$^{d}$ECT$^*$, European Center for Theoretical Studies in Nuclear Physics and 
Related Areas, Strada delle Tabarelle 286, I-38123 Villazzano (TN), Italy \\
\\
E-mail:  \email{eavsar@phys.psu.edu,astasto@phys.psu.edu, dzaslavs@phys.psu.edu}}

\abstract{We investigate the effects of the saturation boundary  on small-$x$ evolution at the next-to-leading order accuracy and beyond. We demonstrate
that the instabilities of the next-to-leading order BFKL evolution are not cured by the presence of the nonlinear saturation effects, and a resummation of the higher order corrections is therefore needed for the nonlinear evolution. 
The renormalization group improved resummed equation in the presence of the saturation boundary is investigated, 
and the corresponding saturation scale is extracted. A significant reduction of the saturation scale is found, 
and we observe that the onset of the saturation corrections is delayed to higher rapidities. This seems to be related to the characteristic feature of the resummed splitting function which at moderately small values of $x$ possesses a minimum.}                        

\begin{document}

\section{Introduction}

It is  well known that the BFKL evolution \cite{Fadin:1975cb,Balitsky:1978ic,Lipatov:1985uk} suffers,
beyond the leading logarithmic order (LL),  from large corrections which are related to the running of the coupling and to kinematical effects, such as energy-momentum conservation.  
Such corrections can be easily added to the leading order formalism in phenomenological applications
but it is also desirable to have a good control of the next-to-leading logarithmic (NLL) corrections in 
the precise theoretical formulation.  BFKL describes an evolution in rapidity where the asymptotic limit 
$s/t \to \infty$ (with $s$ the cms energy and $t$ the momentum transfer)
described by the LL approximation does not correspond to a vanishing coupling strength $\alpha_s$, 
unlike the QCD renormalization group (RG) evolution dictated by the DGLAP equations where the 
asymptotic limit $Q^2\to \infty$ indeed implies that $\alpha_s(Q) \to 0$.  Consequently, the higher 
order corrections in the latter are well controlled in the region of the applicability of the formalism. In the BFKL framework 
on the other hand, as $s$ becomes very large, the evolution receives contributions
from an increasing phase space in momenta where $\alpha_s$ can be very large, and as a consequence 
there is no reason to expect that the higher order corrections can be neglected. Indeed, the calculated 
NLL corrections to the BFKL formalism turn out to be very large \cite{Ciafaloni:1998gs,Camici:1997ij,Camici:1996st,Fadin:1996nw, Kotsky:1998ug,Fadin:1998py}, and  
it is therefore important  to try include the effects of these large corrections in any  application. 

The NLL BFKL evolution is affected by certain problems which lead to unstable results such as negative and oscillating cross sections, see for example \cite{Salam:1999cn}.
These instabilities of the NLL evolution originate from the existence of negative double and triple
poles in the eigenvalue of the evolution kernel.  Certain strategies have therefore been proposed to deal with the 
instabilities of the formalism. These involve the all-order resummation of the dominant parts of the higher order 
corrections, and there exist several prescriptions \cite{Salam:1998tj, Ciafaloni:1998iv, Ciafaloni:1999yw, Ciafaloni:2003rd, Altarelli:1999vw, Altarelli:2001ji, Altarelli:2003hk, Vera:2005jt, White:2006xv,White:2006yh}, all of which are consistent with each other. 
These procedures lead to the so-called ``renormalization group improved'' small-$x$ evolution. 
For a nice and comprehensive review of the NLL formalism, its problems, and the resummation strategies, see \cite{Salam:1999cn}.

When the QCD dynamics is probed at very small-$x$, however, we also expect corrections from 
so-called saturation effects which are related to the formation of strong classical color fields. The effective theory which takes into account these effects is
the Color Glass Condensate (CGC) (for a review see \cite{Iancu:2003xm}). These corrections lead 
to a generalization of the linear evolution equations, which are now instead replaced by a hierarchy of nonlinear equations that go under the name
of the Balitsky-JIMWLK equations \cite{Balitsky:1995ub, JalilianMarian:1997dw,  JalilianMarian:1997gr,  Iancu:2000hn, Ferreiro:2001qy, Iancu:2001ad}.  With certain simplifying assumptions one obtains the compact Balitsky-Kovchegov (BK) equation \cite{Kovchegov:1999yj} which can more easily be used for phenomenology. While the CGC 
formalism has so far only been written down to leading logarithmic order in $x$, in recent years Balitsky and Chirilli have derived the BK equation at the next-to-leading order 
accuracy as well \cite{Balitsky:2008zza, Balitsky:2009yp} (the quark contribution was calculated earlier in 
\cite{Balitsky:2006wa, Kovchegov:2006vj}).  It is therefore hoped that one can thus use these results 
to do phenomenology taking into account both the large NLL corrections to the evolution, and also the nonlinear 
corrections which are expected to be important at small-$x$. Unfortunately, however, the full NLL BK equation is extremely 
complicated and it has so far not been possible to solve it even numerically. 

We shall not present here a numerical solution of the full NLL BK equation, but rather use a very simple, but 
powerful, method to effectively take into account the nonlinear corrections in the full NLL BFKL evolution. 
We will namely solve the full NLL evolution using a so-called saturation boundary which allows us to extract 
the universal properties of the full nonlinear solution such as the energy dependence of the saturation scale, $Q_s$. 
The boundary method was originally used in analytic studies of 
the saturation scale $Q_s$ \cite{Mueller:2002zm, Triantafyllopoulos:2002nz}, and it has since been understood 
that the success and justification of the method can be attributed, at least for a fixed coupling and at asymptotically high energies,  to the formal correspondence between the small-$x$ physics and 
the class of phenomena referred to as reaction-diffusion processes in statistical physics \cite{Munier:2003vc} 
(the evolution equations are said to exhibit ``traveling wave'' solutions). 
In \cite{Avsar:2009pv, Avsar:2009pf}, however, it was explicitly demonstrated via numerical solutions that the 
method works also for a running coupling and for non-asymptotic energies, and see also \cite{Beuf:2010aw} for a 
recent study using the analytic methods of the traveling wave solutions to investigate the universal asymptotic properties
of the small-$x$ evolution beyond the leading order. 

When linearized, the NLL BK equation reduces exactly to the NLL BFKL equation, and the kernel therefore has 
the same eigenvalue as the NLL BFKL one \cite{Balitsky:2008rc, Balitsky:2010jf}. The NLL BK equation thus contains 
the exact same double and triple poles which lead to instabilities in the NLL BFKL evolution. It is 
therefore reasonable to expect that the NLL BK equation will suffer from the same problems which plague the NLL
BFKL evolution. We will  
here demonstrate that the NLL BFKL evolution is indeed unstable also in the presence of saturation effects, and thus 
we can conclude that the nonlinear corrections associated with the physics of saturation, contrary to some earlier hopes, do
not cure the unstable linear NLL evolution. Moreover, we show here that the NLL corrections not associated with the 
running of the coupling (namely those that stem from the nonsingular parts of the DGLAP splitting function and 
the energy scale terms) are extremely important and that they cannot be neglected in any approximation. 

We indeed find a very strong reduction of the saturation scale, $Q_s(x)$, when the full NLL corrections are included. 
The fixed coupling results lead to the plots in figure \ref{fig:qsatfix} where we find that 
$Q_s^2$ is reduced by around two orders of magnitude at $Y=\ln1/x=12$ as compared to the leading order result. 
Actually, the fixed coupling NLL evolution is highly unstable so that one can very well 
expect the full nonlinear evolution to be even more unstable. This makes it rather hard 
to sensibly identify the saturation scale, at least using the widely accepted definitions found in the literature. 
In this case the saturation boundary we apply has a stabilizing effect because the solution in the unstable momentum 
region
is fixed to a certain value imposed by hand. We would certainly not expect 
the full nonlinear evolution to manifest such a regularity. Even in this case, however, we do find that 
the instability eventually kicks in, as $Y$ grows larger, and the solution starts to exhibit non-sensible features. We therefore 
can extract the saturation scale only for a limited interval in $Y$.  It should be mentioned though that we 
in this case have chosen the value $\abar=0.2$ where $\bar{\alpha}_s = \alpha_s N_c/\pi$, with $\alpha_s$ the QCD coupling. 
It is perfectly possible that perhaps for a smaller, but totally unrealistic, value of $\alpha_s$ we could always define a saturation momentum; recall that for $\bar{\alpha}_s \lesssim 0.05$ the Pomeron intercept is positive and real, leading to an exponential growth of the solution. 

Part of the NLL corrections are included in the running of the coupling and when we allow the 
coupling to run, we find a somewhat milder suppression which  is however still very strong, around 
a factor 7 for $Y=12$. We find that the full NLL evolution is extremely sensitive 
on the precise choice of the running of the coupling. While we see that some choices give stable 
and reasonable results, other choices give very unstable results leading to wildly oscillating 
solutions for the transverse momentum distribution obtained from the gluon Green's function. 
Keeping in mind that the differences in the precise choices of the scale of the running 
coupling for the NLL kernel are formally of N$^2$LL 
and N$^3$LL order, we see that the evolution is extremely sensitive to the higher order corrections. 
Moreover it is also very sensitive to the minimum $k=|\vec{k}|$ used in the numerical implementation; lowering this 
minimum beyond some limit causes the solution to exhibit very strong oscillations which make it extremely
unstable.

It therefore seems that some type of resummation procedure as done in the linear case \cite{Salam:1998tj, Ciafaloni:1998iv, Ciafaloni:1999yw, Ciafaloni:2003rd, Altarelli:1999vw, Altarelli:2001ji, Altarelli:2003hk, Vera:2005jt, White:2006xv,White:2006yh} is again 
necessary for stabilizing the evolution.
Such a resummation in the nonlinear case is likely to be a very complicated task which is beyond the scope of the present paper.
We will not present a full solution of the problem here, but instead we take the much simpler approach 
of studying the RG improved evolution in the presence of the saturation boundary.  This was already studied in a semi-analytic way in \cite{Triantafyllopoulos:2002nz}, but only in the asymptotically high energy regime, and therefore potentially important pre-asymptotic corrections might have been missed.
The application of the 
boundary requires some care with the choice of the momentum scales relevant for the 
process under study since the NLL kernel depends on the scale choice. Consequently, the resummation 
procedure also depends on the exact scale choice. 
The saturation boundary explicitly introduces an asymmetry since it
acts as a cut-off on the lower values of the transverse momentum, modifying the linear solution
asymmetrically. Similarly, the BK equation 
describes an asymmetric situation where a rather small probe, such as a virtual photon characterized 
by its virtuality, scatters off a much larger target characterized by a much smaller momentum scale. 
In this paper we consider only the asymmetric situation where a probe with a large scale $Q_A$ scatters
off a target with lower scale $Q_B$ as is the case in Deep Inelastic Scattering (DIS). 

The large $x$ terms present in the RG improved equation have a significant effect on the rapidity evolution.
For the smallest rapidities the evolution is significantly slowed down, and is even negative in some $k$ 
region. This behavior is due to the interplay between leading and non-leading terms which contribute with 
opposite signs. This behavior also manifests itself in the gluon splitting function extracted from the evolution
which exhibits a characteristic ``dip'' when the splitting function is plotted as a function of the longitudinal momentum 
fraction \cite{Ciafaloni:2003kd}. For small and fixed coupling one can do an analytic estimate for the location of this dip, and one finds that it occurs when $Y \sim 1/\sqrt{\bar{\alpha}_s}$. As a result of collinear resummations, this value is, not surprisingly, parametrically far even from the regime where BFKL growth starts to occur, that is from $Y \sim 1/\bar{\alpha}_s$. Thus, the dip should have no consequences for the analytic solution when we enter the BFKL regime and in particular in the asymptotic one, but it is very important for phenomenology since the rapidity window covered by the dip region is non-negligible for realistic values of the total rapidity separation.  
We find for example that the resulting saturation scale $Q_s$ plotted in figure  \ref{fig:qsatresum}, which represents
the main results in this paper,  stays 
constant, fixed by the initial condition, a few units in rapidity $Y$.   
Also, if we plot the resulting distribution in transverse momentum from the RG evolution, we find that saturation plays 
a smaller role in the evolution at these rapidities, as is manifest from the results in figure \ref{fig:nllvsresuminsat01}
where it is seen that the front of the solution essentially progresses with the same speed in both the 
linear and nonlinear cases.

The behavior of the RG improved solution can be compared with that of the pure LL and NLL solutions 
where saturation plays a bigger role. In the LL evolution, supplemented with a running coupling, the nonlinear
corrections are rather important and they significantly reduce the front velocity (\emph{i.e} the rate of 
change with rapidity $Y$ of a point of fixed value for the $k$ distribution). In the NLL case the reduction 
is much smaller but still visible even for phenomenologically relevant values of $Y\lesssim 15$, as
manifest in figure \ref{fig:linvssat_lonlo}.  As mentioned above the difference in the RG improved case 
is on the other hand smaller for the same values of $Y$. 
What this implies for the saturation corrections in the RG improved case is that, as already mentioned, they 
set  in with a delay in rapidity. Needless to say, this behavior has interesting consequences for the search 
for saturation effects in experimental data where the rapidity available is rather limited. To make clear statements 
on the observed phenomenology, however, we would need to do a more careful analysis where the undetermined
parameters and inputs in our approach are set by fitting data. 

The paper is organized as follows.  In the next section we go through the BFKL formalism at both leading and 
next-to-leading order. We describe the choice of the asymmetric scale for the next-to-leading order kernel, 
and we outline the numerical procedure and the extraction of the saturation scale from the numerical solution. 
Then in section \ref{NLOresults} we present the results of our numerical solution for both the fixed and running 
coupling evolutions for the LO and NLO evolutions in the presence of the saturation boundary.  Having demonstrated
the instability of the NLO evolution we then go on to discuss the resummation procedure used in our analysis 
in section \ref{resummedsection}.  We present the exact resummed evolution equation which we solve, and 
we then present the solutions for the saturation scale and the Green's function, demonstrating the suppression 
of the saturation momentum at small values of the rapidity. Finally in section \ref{sec:summary} we briefly summarize
the main findings of our paper.

\section{NLL BFKL with the boundary}
\subsection{General formulation}

\begin{figure}
\begin{center}
\includegraphics[scale=0.5]{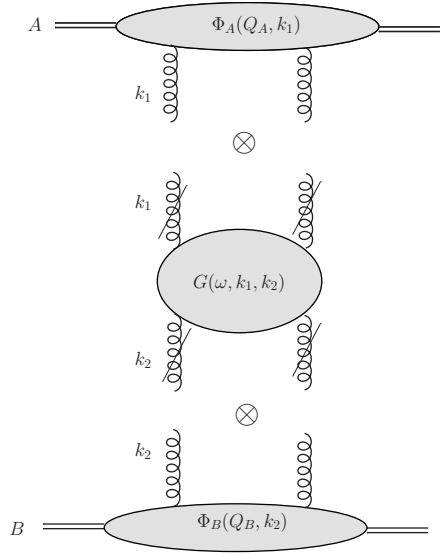}
\caption{\label{fig:figure1} Regge-type factorization formula for the cross section.  The slashed gluon lines indicate 
reggeized gluons.}
\end{center} 
\end{figure}

Let us start this section by recalling the general formulation in QCD of the Regge limit of high energy scattering.
Studies of $\gamma^*\gamma^*$ scattering lead to the formula for the total cross section \cite{Balitsky:1978ic}
(see figure \ref{fig:figure1})
\begin{equation}
\sigma^{AB}(s,Q_A,Q_B) = \int \frac{\dif\omega}{2\pi i} \left( \frac{s}{s_0} \right)^{\omega} \int \frac{\dif^2 \kt_1}{k_1^2} \frac{\dif^2 \kt_2}{k_2^2} \; \Phi_A(Q_A,\kt_1) \; G(\omega; \kt_1,\kt_2) \; \Phi_B(Q_B,\kt_2) \; ,
\label{eq:sigmasym}
\end{equation}
where the functions $\Phi_{A,B}(Q_i,\kt_j)$ are the impact factors for the photons $A$ and $B$ with 
virtualities $Q_A$ and $Q_B$ respectively. The exact 
choice of the scale $s_0$ in the Mellin integral is arbitrary at leading logarithmic order but is important for the 
next-to-leading order calculation. The function $G(\omega; \kt_1,\kt_2)$ is referred to as the ``BFKL Green's function'' 
(or ``gluon Green's function''), and should be thought of as the gauge invariant generalization of the vacuum expectation
value of four off-shell gluons. It satisfies the BFKL equation, which can be written (in the case of forward scattering)
\cite{Fadin:1975cb,Balitsky:1978ic,Lipatov:1985uk}
\begin{equation}
\omega \, G(\omega;\kt,\kto) \; = \; \delta^2(\kt-\kto) \; + \;\int \frac{\dif^2 \ktp}{\pi^2} \,  K(\kt,\ktp) \,G(\omega;\ktp,\kto) \; ,
\label{bfklgreen}
\end{equation}
where the kernel of the equation is known up  the next-to-leading logarithmic (NLL) order in 
$\ln 1/x$  \cite{Camici:1997sh,Ciafaloni:1998gs,Camici:1997ij,Camici:1996fr,Camici:1996st,Fadin:1996nw,Fadin:1996zv,Kotsky:1998ug,Fadin:1998py}
\begin{equation}
K(\kt_1,\kt_2) \; = \; K_0(\kt_1,\kt_2) \; + \; K_1(\kt_1,\kt_2) \; + {\cal O}(\alpha^3(\mu^2)) \; .
\label{bfklkernel}
\end{equation}
It is here understood that $K_0$ and $K_1$ are of order $\alpha_s$ and $\alpha_s^2$ respectively 
(as clear from equations \eqref{eq:llbfkl} and \eqref{eq:nllbfkl} below). 

We shall be  solving the BFKL equation using both kernels $K_0$ and $K_1$. In addition we will be interested 
in studying the nonlinear evolution equation obtained after introducing a saturation boundary which modifies 
the action of the linear kernel $K$. We will explain this procedure further below. Let us mention that we will 
generally be solving the BFKL equation with a generic initial condition in $\kt$. This corresponds to defining a new 
function $\mathcal{F}(\omega, \kt)$ by
\begin{equation}
\mathcal{F}(\omega,\kt;Q_B) \equiv \int \frac{\dif^2 \kt_2}{k_2^2}G(\omega; \kt,\kt_2) \; \Phi_B(Q_B,\kt_2)\; ,
\end{equation}
which then satisfies the equation 
\begin{equation}
\omega \mathcal{F}(\omega,\kt;Q_B) = \frac{\Phi_B(Q_B,\kt)}{k^2}+ \;\int \frac{\dif^2 \ktp}{\pi^2} \,  K(\kt,\ktp)
\mathcal{F}(\omega,\ktp;Q_B)\; .
\end{equation}
This implies that we may write the cross section \eqref{eq:sigmasym} as
\begin{equation}
\sigma^{AB}(s,Q_A,Q_B) = \int \frac{\dif\omega}{2\pi i} \left( \frac{s}{s_0} \right)^{\omega} \int \frac{\dif^2 \kt_1}{k_1^2}
\Phi_A(Q_A,\kt_1) \mathcal{F}(\omega,\kt_1;Q_B)\; .
\label{sigma2}
\end{equation}
In the following we shall keep the dependence of $\mathcal{F}$ on $Q_B$ implicit.

Let us now turn to the explicit expression for the BFKL kernel up to the next-to-leading order. 
We consider the solutions which are averaged over the angle. 
The leading logarithmic (LL) order kernel (after the angular averaging) is given by  
\begin{equation}
\int \frac{\dif k^2_2}{\pi} \, K_0(k_1,k_2) \, f\bigl(k_2^2\bigr)  =   \abar(\mu^2)  \int \dif k_2^2  \frac{1}{|k_1^2-k_2^2|} \biggl( f\bigl(k_2^2\bigr)-2\, \frac{{ \min}\bigl(k_1^2,k_2^2\bigr)}{k_1^2+k_2^2}f\bigl(k_1^2\bigr) \biggr)   \; , 
\label{eq:llbfkl}
\end{equation}
where we remind once again that $\abar \equiv \alpha_sN_c/\pi$. 
In expressing the explicit form of the action of the kernel, we have above introduced an auxiliary function $f$ to simplify the notation.
We also use $\mu$ to generically denote the scale of the strong coupling.  Note that the leading order kernel 
does not have any dependence on $\mu$ since any difference in the choice of scale is formally of next-to-leading 
order. The leading order kernel is therefore scale-invariant. Physically this can be interpreted as the fact that 
in the leading order one is taking the limit of infinite collision energy which implies that any 
other finite scale in the process (for example transverse momenta, masses, etc) can be 
neglected. The next-to-leading order kernel is instead given by
\begin{multline}
\int \frac{\dif k^2_2}{\pi} \, K_1(k_1,k_2) \, f\bigl(k_2^2\bigr)  = 
-\frac{1}{4} \abar^2(\mu^2) \int \dif k_2^2  \Biggl\{\frac{1}{|k_1^2-k_2^2|} \biggl( f\bigl(k_2^2\bigr) - \frac{2\min\bigl(k_1^2,k_2^2\bigr)}{k_1^2+k_2^2}f\bigl(k_1^2\bigr) \biggr) \hspace{5em}\\
\hspace{5em}\times\biggl[ \biggl(\frac{11}{3}-\frac{2n_f}{3N_c}\biggr)\ln \frac{|k_1^2-k_2^2|^2}{\mu^2\,\max(k_1^2,k_2^2)} - \biggl(\frac{67}{9}-\frac{\pi^2}{3}-\frac{10}{9}\frac{n_f}{N_c}\biggr) \biggr] \\
-f\bigl(k_2^2\bigr) \Biggl[\frac{1}{32} \biggl(1+\frac{n_f}{N_c^3}\biggr)\Biggl(\frac{2}{k_1^2}+\frac{2}{k_2^2}+\biggl(\frac{1}{k_2^2}-\frac{1}{k_1^2}\biggr)\ln \frac{k_1^2}{k_2^2}\Biggr) + \frac{1}{|k_1^2-k_2^2|} \Biggl( \ln \frac{k_1^2}{k_2^2} \Biggr)^2 \\
+ \Biggl( 3 + \biggl(1+\frac{n_f}{N_c^3}\biggr) \biggl(\frac{3}{4} - \frac{(k_1^2+k_2^2)^2}{32 k_1^2 k_2^2} \biggr)\Biggr) \int_0^{\infty} \frac{\dif x}{k_1^2 + x^2 k_2^2} \ln \biggl|\frac{1+x}{1-x}\biggr| \\
-\frac{1}{k_1^2+k_2^2} \Biggl(\frac{\pi^2}{3} + 4 \Li\biggl[\min\biggl(\frac{k_1^2}{k_2^2},\frac{k_2^2} {k_1^2}\biggr)\biggr]\Biggr)\Biggr]\Biggr\} 
+\frac{1}{4}  \abar^2(\mu^2) \Biggl( 6 \zeta(3) - \frac{5 \pi^2}{12} \biggl( \frac{11}{3}-\frac{2 n_f}{3N_c} \biggr) \Biggr) f(k_1^2) \; .
\label{eq:nllbfkl}
\end{multline}
Here $n_f$ is the number of quark flavors, and $\Li$ is the dilogarithm function.
The scale dependence on $\mu$ in the expression for $K_1$ is related to  the running of the QCD 
coupling.   Starting from the NLL order the kernel $K(\kt_1,\kt_2;\mu)$ is thus no longer scale-invariant. The above 
form of the kernel was obtained for the so-called symmetric scale choice, see for example \cite{Ciafaloni:1998gs}.
 This means that the solution $G(\omega;k,k_0)$ to the BFKL equation with the NLL kernel above should be used for 
the computation of the cross section in \eqref{eq:sigmasym} with $s_0=Q_A \,Q_B$. A physical example of
this case is given in $\gamma^*\gamma^*$ scattering where the virtualities of the photons are comparable.
On the other hand for DIS, still using the same formula for the cross section  \eqref{eq:sigmasym}, 
the scale $Q_A$ is the virtuality of the photon and $Q_B$ is the scale 
characterizing the hadron target. Therefore for this situation  $Q_A\gg Q_B$, 
and the choice of  scales will be asymmetric, i.e. $s_0=Q_A^2$.

\subsection{Scale choice in the presence of the saturation boundary}

As will be demonstrated  in the numerical analysis,  the scale choice in the presence of the saturation boundary leads 
to sizeable variations in the form of the solution. This is so because the scale choice alters the form of 
the kernel \eqref{eq:nllbfkl}. In the Mellin space, it is linked to the fact that the scale choice changes the terms which contain 
triple collinear poles \cite{Ciafaloni:1998gs,Salam:1998tj}. When combined with the nonlinear evolution this can lead to 
sizable differences due to the fact that the boundary is also asymmetric with respect to the infrared and ultraviolet 
regions (when impact parameter is not taken into account). To be precise, in the case of the translationally invariant BK 
equation the effect of the nonlinear term is such that it cuts off the  infrared region of momenta.  On the other hand,  in 
writing down the explicit form of the NLL kernel \eqref{eq:nllbfkl}, it is implicitly assumed that the evolution is 
symmetric with respect to the two momentum scales, $k_1 \leftrightarrow k_2$, which is not the case of the BK evolution. Therefore the correct treatment of the energy scale 
choice is linked with the problem of the symmetry of the evolution with respect to the target and projectile. This problem 
is rather difficult and it is plausible that for the complete solution one needs to take into account the other contributions, 
like Pomeron loops, which will guarantee the symmetry of the evolution, 
\cite{Mueller:2004sea,Iancu:2004iy,Triantafyllopoulos:2005cn}.
We are not going to address this important, and difficult, issue here but rather pick a scale choice that is relevant for the DIS process 
off a nucleus for which the BK evolution is supposed to be the correct treatment.
Therefore we adopt the choice,  $s_0 = Q^2$,  which is relevant  for the DIS process where $Q \gg Q_0$, 
with being $Q$ the virtuality of the photon whereas $Q_0$ characterizes the target nucleus or a proton. Consequently 
one needs to perform the corresponding scale change in the BFKL kernel above, which amounts to shifting
the characteristic function at the NLL level \cite{Fadin:1998py},

\begin{equation}
\widetilde{\delta}(\gamma) \; = \; \chi_1(\gamma) -2 \chi_0(\gamma) \chi_0^{\prime}(\gamma) \; .
\label{Mellinshift}
\end{equation}
The functions in Mellin space are here defined as
\begin{equation}
\abar \,\chi_0(\gamma)=\int \frac{\dif k^2_2}{\pi} \,  K_0(k_1,k_2) \left(\frac{k_2^2}{k_1^2} \right)^{\gamma-1} \;  , \;
\label{kernelshift}
\end{equation}
and
\begin{equation}
\abar^2 \,\chi_1(\gamma)=\int \frac{\dif k^2_2}{\pi} \, K_1(k_1,k_2) \left(\frac{k_2^2}{k_1^2} \right)^{\gamma-1} \;  . \;
\label{kernelshift2}
\end{equation}
The explicit expressions for the characteristic function are
\begin{equation}
\label{eq:llchi}
 \chi_0(\gamma) = 2\psi(1)-\psi(\gamma)-\psi(1-\gamma) \; , \; \; \;  \psi \equiv \frac{1}{\Gamma}\frac{\dif\Gamma(\gamma)}{\dif\gamma}
\end{equation}
at the level of the LL approximation, and
\begin{align}\nonumber
 \chi_1(\gamma) &= -\frac{b}{2} \bigl[\chi^2_0(\gamma) + \chi'_0(\gamma)\bigr]
  -\frac{1}{4} \chi_0''(\gamma)
  -\frac{1}{4} \left(\frac{\pi}{\sin \pi \gamma} \right)^2
  \frac{\cos \pi \gamma}{3 (1-2\gamma)}
  \left(11+\frac{\gamma (1-\gamma )}{(1+2\gamma)(3-2\gamma)}\right) \\
 &\quad +\left(\frac{67}{36}-\frac{\pi^2}{12} \right) \chi_0(\gamma)
  +\frac{3}{2} \zeta(3) + \frac{\pi^3}{4\sin \pi\gamma}  \nonumber\\
 &\quad - \sum_{n=0}^{\infty} (-1)^n
 \left[ \frac{\psi(n+1+\gamma)-\psi(1)}{(n+\gamma)^2}
 +\frac{\psi(n+2-\gamma)-\psi(1)}{(n+1-\gamma)^2} \right] \;
\label{eq:nll}
\end{align}
at the NLL level. 
In the above equation we have set $n_f=0$ and this will be our assumption for the rest of this paper. Obviously, for phenomenological applications the quarks should be included in the evolution. For the purpose of this work, however, we will assume that the dynamics is constrained to the gluon sector only. 

The corresponding momentum space representation of the shift  part of the kernel is expressed as 
\begin{multline}
I_{\text{shift}}=2\int_0^1 \frac{\dif u}{1-u} f(u) \biggl[\frac{1}{2} \ln^2 u - 2 \ln u \ln(1-u)\biggr] \; + \\
+ \; 2\int_1^{\infty}\frac{\dif u}{u-1}f(u) \biggl[-\frac{1}{2} \ln^2 u -2 \ln u \ln\biggl(1-\frac{1}{u}\biggr) \biggr] \; .
\end{multline}
It can be readily checked that the above equation with  
\begin{equation}
f(u) = u^{\gamma-1} \; ,
\end{equation}
gives 
\begin{equation}
I_{\text{shift}} = - 2\chi_0(\gamma) \chi_0^{\prime}(\gamma) \; .
\end{equation}

\subsection{Numerical implementation}

For the numerical implementation, the explicit form of the kernel (\ref{eq:nllbfkl}) is not very suitable 
due to the large cancellations between the terms which constitute the 
contribution to the nonsingular part of the DGLAP splitting function. In particular there are large 
superleading logarithms which cancel between terms in third and fourth lines of (\ref{eq:nllbfkl}).
To simplify the numerical procedure, and to obtain an accurate solution, we can instead rewrite 
these terms in a suitable way.  We start by using the following form of the integral
\begin{equation}
\int_0^{\infty} \frac{\dif x}{k_1^2 + x^2 k_2^2} \ln \biggl|\frac{1+x}{1-x}\biggr|  
\; = \;  \frac{1}{k_1 k_2} \biggl[ \ln \frac{k_{>}^2}{k_{<}^2} \tan^{-1} \frac{k_{<}}{k_>} + 2 {\Im} \Li \biggl(i \frac{k_<}{k_>}\biggr) \biggr] \; , 
\end{equation}
where $k_{<(>)}= \min(\max)(k_1,k_2)$ to express the difficult parts of the kernel (\ref{eq:nllbfkl}) in the following form
\begin{multline}
 \frac{1}{32} \Biggl( \frac{2}{k_1^2}+\frac{2}{k_2^2}+\biggl(\frac{1}{k_2^2}-\frac{1}{k_1^2}\biggl)\ln \frac{k_1^2}{k_2^2}\Biggr) \\
+ \Biggl(3 + \biggl( \frac{3}{4} - \frac{(k_1^2+k_2^2)^2}{32 k_1^2 k_2^2} \biggr)\Biggr) \int_0^{\infty} \frac{\dif x}{k_1^2 + x^2 k_2^2} \ln \biggl|\frac{1+x}{1-x}\biggr| \\
= \; \frac{1}{32} \biggl( \frac{2}{k_>^2} -\frac{1}{k_>^2} \ln \frac{k_>^2}{k_<^2} \biggr) 
+ \; \frac{1}{k_> k_<} \biggl[ 3+ \frac{1}{32} \biggl(22-\frac{k_<^2}{k_>^2}\biggr)\biggr] \biggl[ \frac{k_<}{k_>} \ln \frac{k_>^2}{k_<^2} +2 \frac{k_<}{k_>} \biggr] \\
+ \; \frac{1}{k_> k_<}\biggl[ 3 + \frac{22}{32} -\frac{1}{32} \biggl( \frac{k_<^2}{k_>^2} + \frac{k_>^2}{k_<^2} \biggr) \biggr] \biggl[ \ln \frac{k_>^2}{k_<^2} \, \operatorname{S_1}\biggl(\frac{k_<}{k_>}\biggr)+2 \, \operatorname{S_2}\biggl(\frac{k_<}{k_>}\biggr) \biggr] \; .
\label{eq:expression}
\end{multline}
The functions $\operatorname{S_1}$ and $\operatorname{S_2}$ are series expansions of $(\tan^{-1}(x)-x)$ and $(\Im \Li(x)-x)$ respectively. We have checked that it is sufficient to retain only around twenty terms in the expansions to get accurate results.


\subsection{Saturation scale from the boundary method} 

One of the main objectives in this paper is to extract the saturation scale $Q_s$ from the NLL and RG improved 
evolutions using the boundary method. In this section we therefore first describe the definition of the saturation 
scale used, and having done that we then describe the precise numerical method in which the definition is 
employed to obtain $Q_s$. 

\subsubsection{The definition of the saturation scale}

The usual definition of the saturation scale follows from the solution to the nonlinear 
BK equation. The object satisfying the BK equation is the coordinate space scattering amplitude\footnote{
Equivalently the BK equation can also be written for the dipole ``S-matrix'' defined as 
$S=1-\mcN$, see equations \eqref{BK} and \eqref{NLOBK}.}
$\mcN(\hat{s},r)$ for a dipole of size $r$. Here we denote $s/s_0$ by $\hat{s}$. 
In this case one can define the saturation momentum $Q_s$ as the scale which separates the 
regions where $\mcN$ is ``small'' and thus follows a linear evolution, 
and where it is nearly saturated at $\mcN=1$ and its evolution is completely nonlinear. The exact definition
of $Q_s$ is always somewhat ambiguous since it depends on the precise value of $\mcN$ where one 
chooses to separate the linear 
and the nonlinear regions. 

In determining $Q_s$ one defines first the critical dipole size $r_s(\hat{s})$ by 
$\mcN(\hat{s},r_s(\hat{s})) = c < 1$ where $c$ is a given constant smaller than 1. Then one can take $Q_s=C \,r_s^{-1}$ 
where $C$ is another constant. In the Golec-Biernat--Wusthoff model  \cite{Golec-Biernat:1999qd,Golec-Biernat:1998js} for instance, 
one chooses $C=2$. 
The constant $c$ can be chosen to be around $0.1\, - \, 0.5$, the exact value will determine only 
the normalization of $Q_s$ which is never under full control theoretically. What is determined by the 
perturbative evolution is the scale dependence of $Q_s$, i.e. the $\hat{s}$ dependence. 
Note that the saturation of $\mcN$ does not necessarily imply the saturation of 
the cross section $\sigma(\hat{s},r)$ of the dipole. The saturation of the cross section is a nonperturbative problem 
which is not solved by the perturbative nonlinear evolution equation for $\mcN$. 
To obtain the cross section $\sigma$ one needs to integrate $\mcN$
over the impact parameter $b$
\begin{equation}
\sigma_\text{dip}(\hat{s},r) = 2 \int \dif^2 b \, \mcN(\hat{s},r, b) \; .
\label{sigmadip}
\end{equation}
Obviously the behavior of $\mcN$ at large $b$ is nonperturbative and needs to be modeled phenomenologically. 
In the GBW model the dipole cross section $\sigma$ is taken to be 
\begin{equation}
\sigma_\text{dip}(x,r) = \sigma_0 \Biggl ( 1 - \exp\biggl(-\frac{r^2}{4 \, r^2_s(x)} \biggr) \Biggr) \; .
\label{gbw}
\end{equation}
Here one sets $\hat{s} = x^{-1}$. In this case $r_s$ is the length scale above which the cross section 
of the dipole saturates to the constant $\sigma_0$. If one assumes that the $b$ dependence factorizes 
as $\mcN(r,b) = \mcN(r) \, S(b)$ then $\sigma_0=2\int \dif^2b \, S(b)$, and in \eqref{gbw} clearly $r=r_s(x)$
is a line of constant $\mcN$. Of course the assumption that $\sigma_0$ does not evolve with $x$ 
implies in this case the complete saturation of the total cross section. 
It is well known that the $x$ dependence of the fitted $r_s(x)$ 
agrees in form with the one obtained from the solution to the BK equation, using the definition of constant $\mcN$, 
but also that the leading order evolution gives a much too steep growth with $1/x$.  

In this paper we are, however, solving the momentum space BFKL equation using the saturation boundary. 
The question then is how exactly we should define the saturation scale from the solution to the evolution equation.
It is  possible to consider different choices. For example, in \cite{Kutak:2004ym} $Q_s$ was defined as a line along which the difference between the linear and the nonlinear solutions to $\mcF$ (in that case obtained from the leading order BK equation)
is  of a certain magnitude. This, however, gives an energy dependence somewhat different than the coordinate space definition. 

We shall therefore choose a simpler prescription whereby  just as in the coordinate space (similarly to \eqref{gbw}) we define $Q_s$ via 
\begin{equation}
\mcF(\hat{s},Q_s(\hat{s})) = \mathrm{const} \times \sigma_0\; .
\label{qsatdef}
\end{equation}
Notice that according to \eqref{sigma2}, $\mathcal{F}$ equals the cross section when the impact 
factor $\Phi_A/k^2$ is a delta function in the transverse momentum. That would be the case if the scattering object $A$ itself 
is a parton. Thus one can think of $\mathcal{F}$ as the cross section of a parton impinging 
on the target particle. Therefore the definition \eqref{qsatdef} bears a certain resemblance to the definition \eqref{gbw}.
Another partial motivation for this simple choice comes from the fact that this definition gives an $x$ dependence of   
$Q_s$ which is rather consistent with the empirical dependence extracted from data. We are here 
not concerned with the exact value of $\sigma_0$ which we shall leave unspecified\footnote{As we are here not doing 
any phenomenological fit this will not be important. The exact value $\sigma_0$ would be determined as a
fit parameter in any practical application.}. The exact value of the right hand side of \eqref{qsatdef}
will determine only the absolute normalization of $Q_s$. 

Another motivation that can be given for the definition \eqref{qsatdef} directly in momentum space is, at least quasi-classically, a number density in the transverse phase space. 
In the Color Glass Condensate model, the equivalent number density
\begin{equation}
\frac{\dif N}{\dif Y\dif^2k \dif^2b} \; ,
\end{equation}
of the 
classical fields saturates when it is of order $1/\alpha_s$ (see for example \cite{Iancu:2002xk}), 
and it leads to a definition of the saturation scale exactly as in  \eqref{qsatdef} when integrated over 
the impact parameter.  
Now, our $\mcF$, which satisfies the BFKL equation and which can be used in a formula like \eqref{sigma2} 
to calculate the total cross section, cannot literally be thought of as a number density in the phase space. 
It rather corresponds to a cross section at the parton level which is conceptually a different object than a 
phase space number density, since the gluon Green's function does not have an operator definition which would match exactly the definition of a number density. 

The object defined as a phase space number density, which is essentially the expectation value of the 
field-strength tensor $\langle F^{+i}F^{+i}\rangle$ in 
light-cone gauge (see equation (2.15) in \cite{Iancu:2002xk}) does, however, also satisfy the BFKL equation 
at the linear level (see (3.59) in \cite{Iancu:2002xk}). On the other hand it does not satisfy the BK equation 
at the nonlinear level, but some equation more complicated than BK (this equation can be 
obtained by applying the JIMWLK kernel to the operator definition of the quasi-classical number 
density given by equation (2.18) in \cite{Iancu:2002xk}). The point now 
is that it does not matter for our analysis what the exact nonlinear equation is, no matter how complicated 
it may be. The boundary method is a generic method which can extract the universal properties of the full
solution, such as the $\hat{s}$ dependence of $Q_s$, and it therefore works for nonlinear equations 
which are in the universality class of the BK equation, that is equations whose linear parts are driven 
by the BFKL kernel.  Therefore if we simply regard $\mathcal{F}$ only by its property that it satisfies the 
BFKL equation, that is to say if we simply forget the would be operator definition of the gluon Green's function
and only consider the equation it obeys, 
then we do not have to care whether it actually corresponds to the phase space number density or 
to the dipole scattering cross section, we will obtain the same solution for the saturation scale. Since the definition
\eqref{qsatdef} works in both cases we conclude that it is indeed the optimal choice for our problem. 

We therefore now move on to the exact implementation of the boundary which effectively takes into 
account the missing nonlinear effects.

\subsubsection{Application of the saturation boundary}

In accordance with the definition \eqref{qsatdef}, we shall apply the saturation boundary in our 
numerical treatment as follows.  First of all, as explained in the previous section we need not specify the exact 
value of $\sigma_0$ for the purposes of our study. Let us therefore define the new function $F(\hat{s},k)$ 
by
\begin{equation}
F(\hat{s},k) = \frac{\mathcal{F}(\hat{s},k)}{\sigma_0} \; ,
\end{equation}
which means that the condition \eqref{qsatdef} defining the saturation scale now simply reads
\begin{equation}
F(\hat{s},k=Q_s) = \text{const.}
\label{qsatdef2}
\end{equation}
Obviously $F$ satisfies the exact same equation as $\mathcal{F}$.

We then define first a so-called ``critical value'', $c$, which for convenience will be taken as a number of order 1. This 
number can be close to the constant in \eqref{qsatdef} but ideally it should be slightly smaller\footnote{This is 
so since our $c$ is loosely speaking the value where the effects of saturation first starts to play a role 
while $Q_s$ can be thought of the scale below which the evolution is really dominated by the nonlinearities.}. 
At each step in the numerical solution of $F$ we then define the corresponding critical transverse
momentum scale $k_{c}$ via
\begin{equation}
F\bigl(\hat{s},k_{c}(\hat{s})\bigr) = c \; . 
\label{critvalue}
\end{equation}
The boundary will now be applied to those transverse momenta which are below the critical scale by some 
magnitude determined by a second parameter $\Delta$. More precisely, if  by $\rho$ we denote 
the logarithmic scale $\rho \equiv \ln (k^2/k^2_0)$ then $F(\hat{s},\rho)$ is forced to satisfy the 
given boundary condition for all 
\begin{equation}
\rho \leq \rho_c - \Delta, \,\,\,\,\,\, \rho_c \equiv \ln\bigl( k^2_{c}(\hat{s})/k^2_0\bigr) \; .
\end{equation}
We shall choose the arbitrary $k_0$ as the minimum $k$ of our numerical computation.  

In the original analysis of the method in \cite{Mueller:2002zm} the boundary was chosen to be 
totally absorptive, i.e. 
\begin{equation}
F(\hat{s},\rho) = 0 \,\,\, \mathrm{for} \,\, \mathrm{all}  \,\,\, \rho \leq \rho_c(\hat{s}) - \Delta \; .
\label{absorpbound}
\end{equation}
This is the choice most appropriate for the analytic analysis since, 
after the leading exponential behavior has been factored out, the problem can then 
be formulated as a random walk in the presence of an absorptive wall \cite{Mueller:2002zm}. 
For the numerical solution on the other hand we can chose any condition which cuts off the 
power-like growth of the linear evolution with  $\hat{s}$. 
Actually the totally absorptive boundary has 
to be applied with some care, in the numerical treatment, when the evolution is not "fast enough".  
What happens in this case 
is that when all the contribution below the boundary $ \rho_c(\hat{s}) - \Delta$ is cut completely by hand, 
the next step of the evolution may not be strong enough to push the solution above the critical value $c$.
If this happens, then at this next step the boundary does not get applied by definition, since the solution 
completely falls below $c$. As saturation then suddenly switches off, there is an accumulation of the 
solution just around the boundary which can give a "spike" in the solution (that is the solution suddenly jumps 
high above the critical value before it is set to zero), and moreover some of the 
previous points in $k$ where the solution was set to zero may now grow which in turn implies that the front may 
actually, in a single step of the evolution, move 
in the wrong direction (to smaller $k$). We have observed this somewhat peculiar behavior also in the 
case of the CCFM evolution equation where the evolution is suppressed at large $k$ due to the restriction 
imposed by angular ordering.  It also appears to be the case in the RG improved evolution we study 
below. On the other the leading order BFKL evolution always seems fast enough 
so that this problem never appears. We do not find that this is a major obstacle for the numerical implementation 
of the totally absorptive boundary, but it does require some care in the precise treatment. 
We therefore also consider a second boundary which is more straightforward to implement numerically. 
In this case we simply freeze $F$ at the boundary,  and one can show that the two implementations are equivalent in the asymptotic regime. That is, at each step in the solution 
we let
\begin{equation}
F(\hat{s},\rho) = F(\hat{s},\rho_c -\Delta)\,\,\, \mathrm{for} \,\, \mathrm{all}  \,\,\, \rho \leq \rho_c(\hat{s}) - \Delta \; .
\label{frozenbound}
\end{equation}

Let us here emphasize that these choices should not be interpreted literally as to how exactly 
saturation would act on $F$. For example, freezing $F$ at a constant value does not imply that 
one should think of the dynamics as representing the physical result of the saturation of some gluon occupation number 
at a fixed value. The whole idea of the boundary 
method is that it does not matter how  the linear growth is cut off.  We have chosen these two boundaries 
for their simplicity, not because they would represent a more accurate representation of the true nonlinear 
terms compared to other possible choices. 

Studying the evolution in the presence of these two boundary conditions we shall extract the universal 
properties of the solution that are independent of the precise conditions. As is evident from our construction, 
$c$ and $\Delta$ can be thought of as free parameters related to the freedom in 
choosing the precise way via which the missing nonlinear terms are accounted for. 
In reality, however, these parameters are not completely free. 
First, they are correlated as  $\Delta \sim \ln1/c$ \cite{Mueller:2002zm}. Secondly, the value of $c$ in 
\eqref{critvalue} cannot be completely arbitrary.  As we noted above, $c$ should be of the same order 
of the constant used in \eqref{qsatdef}, and the latter is of order 1.  We have chosen our default value 
to be $c=0.4$, but as in the previous works  \cite{Avsar:2009pv, Avsar:2009pf, Avsar:2010ia} we have also 
considered  the possible sensitivity of the solution to different choices of $c$ (and $\Delta$).

\section{Results for NLL BFKL with and without saturation}
\label{NLOresults}

\subsection{Results with fixed coupling}

\begin{figure}
\begin{center}
\includegraphics[scale=1.2]{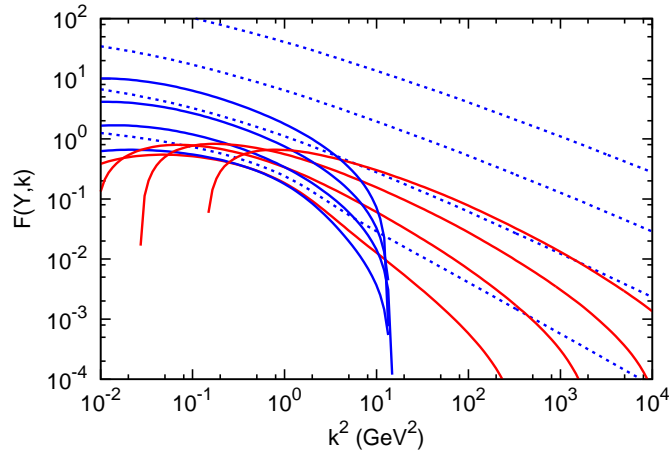}
\caption{\label{fig:fixnlllin} Solutions to the leading and next-to-leading BFKL equations for a fixed coupling 
$\abar = 0.2$ and for $Y=2,6,10,14$: NLL BFKL with asymmetric scale choice (solid red), NLL BFKL with symmetric scale 
choice (solid blue), LL BFKL (dotted blue). }
\end{center} 
\end{figure}

In this section we present the results obtained using a fixed coupling, $\abar = 0.2$. 
We start by studying the linear evolution equations. 
For the results shown in figure \ref{fig:fixnlllin}, and in subsequent plots, we have 
chosen the initial condition
\begin{equation}
F_0(k) = c \cdot \exp \biggl(-\frac{k^2}{k_{in}^2} \biggr) \; ,
\label{initialdistrb}
\end{equation}
with $k_{in} = 1$ GeV. Here $c$ is the parameter in \eqref{critvalue} which determines the 
critical value beyond which the boundary is applied to $F$.  We also use the logarithmic variable
$Y = \ln \hat{s} = \ln s/Q^2$ to denote the energy dependence of the solution. 
The comparison between 
the LL and the NLL BFKL solutions is shown in figure \ref{fig:fixnlllin}. Here the NLL 
solutions are shown for both the symmetric and the asymmetric scale choices.  As well 
known, see for example \cite{Fadin:1998py,Andersen:2003wy,Andersen:2003an,Ciafaloni:2003rd},  the NLL evolution is significantly slower than the leading order evolution. We 
also clearly see the effects of the negative Mellin space poles in the collinear and anti-collinear limits. 
For the symmetric solution (the solid blue curves in figure \ref{fig:fixnlllin}) 
the collinear pole is clearly visible in the plot as the solutions turn rapidly negative  
at moderately high values of the momentum $k$. For the asymmetric scale 
choice the shift in the characteristic function \eqref{Mellinshift} removes the collinear triple pole
while there is still the double pole. As a consequence it turns negative ``later'' (i.e. at higher 
values of  $k$) than the symmetric solution. On the other hand 
the pole at the anti-collinear end causes the solution to turn negative at the smaller $k$ 
values as clearly visible in the figure.  

The apparent instability of the solution suggests that the full-linear solution might very well be 
even more unstable. The precise behavior will of course depend on the exact shape of the nonlinear
terms and how these act, but it is clear that the existence of a nonlinear damping term $F^2$ or
$F^3$ can potentially cause further instabilities when $F$ turns negative. It is therefore not all clear whether 
one would obtain any sensible results.  Of course one can imagine that the results are stable for a smaller 
coupling. For example it is known that for $\bar{\alpha}_s \lesssim 0.05$ the Pomeron intercept is positive and real
in the NLL case, and this leads to an exponential growth of the solution. Such small values of $\alpha_s$ 
are of course hardly realistic.

The boundary method on the other hand is very stable by construction
since the solution below the boundary is set to a fixed value by hand. Despite this, however, 
we will see below that the final result is nevertheless unstable.  Note also that, the successful implementation 
of the boundary method should be independent of the precise value of the critical value $c$. 
For the LL solution in figure \ref{fig:fixnlllin} it is clear that no matter what $c$ is chosen, the linear solution 
eventually reaches this point, and the saturation boundary therefore gets implemented. The 
difference between various values of $c$ is simply in the normalization of $Q_s$ which of 
course is a measure of the strength of saturation (and therefore of the value of $c$). For the 
asymmetric NLL solution (solid red lines) in figure \ref{fig:fixnlllin}, it is, however, clear that the peak of 
the solution is bounded due to the solution turning negative at smaller $k$. In this case 
if we would choose $c$ to be say larger than 1, then it appears that the boundary would never get
applied, and consequently it would have no effect whatsoever on the linear solution. Clearly this is 
a peculiarity of the unstable NLL evolution, and the results for the fixed coupling case should therefore not be taken too seriously.  What we can learn from the fixed coupling case is on the other hand that the full non linear solution 
might very well not give any sensible results (it would of course be highly desirable if this can be checked explicitly).

\begin{figure}[t]
\begin{center}
\includegraphics[angle=270, scale=0.4]{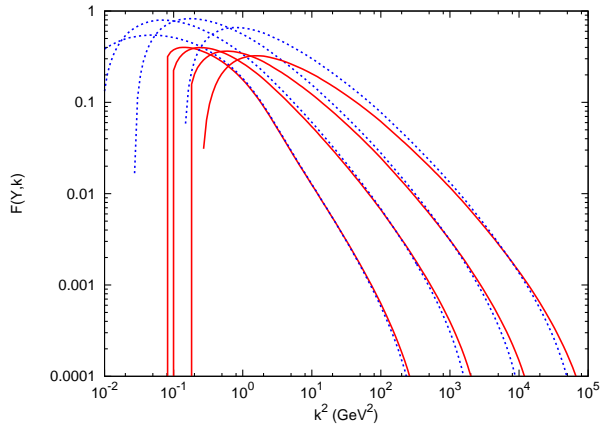}
\end{center} 
\caption{\label{fig:fixnlllinabs} Comparison of the NLL BFKL solution with (solid red) and without (dotted blue)
the absorptive saturation boundary \protect \eqref{absorpbound} for fixed 
$\abar = 0.2$, and $Y=2,6,10,14$.}
\end{figure}

Keeping these points in mind we now apply the saturation boundary to the linear solutions shown in 
figure \ref{fig:fixnlllin}.
In accordance with the discussion above we study the nonlinear evolution at the next-to-leading 
order only with the asymmetric scale choice obtained by applying the shifts in \eqref{Mellinshift}, 
\eqref{kernelshift} and \eqref{kernelshift2}.  
In figure 
\ref{fig:fixnlllinabs} we compare the fixed coupling NLL solutions with and without the absorptive saturation boundary.   
We see 
that the instability of the NLL solution at lower $k$ is not removed completely by the condition of the absorptive
boundary. For the solution at $Y=14$ we see that $F$ turns negative before it is set to 0 by the  
boundary condition. As in the linear evolution the solution also turns negative at higher $k$ which is of course 
expected since the nonlinearities do not cure the unstable high-$k$ behavior. 

\begin{figure}[t]
\begin{center}
\includegraphics[angle=270, scale=0.35]{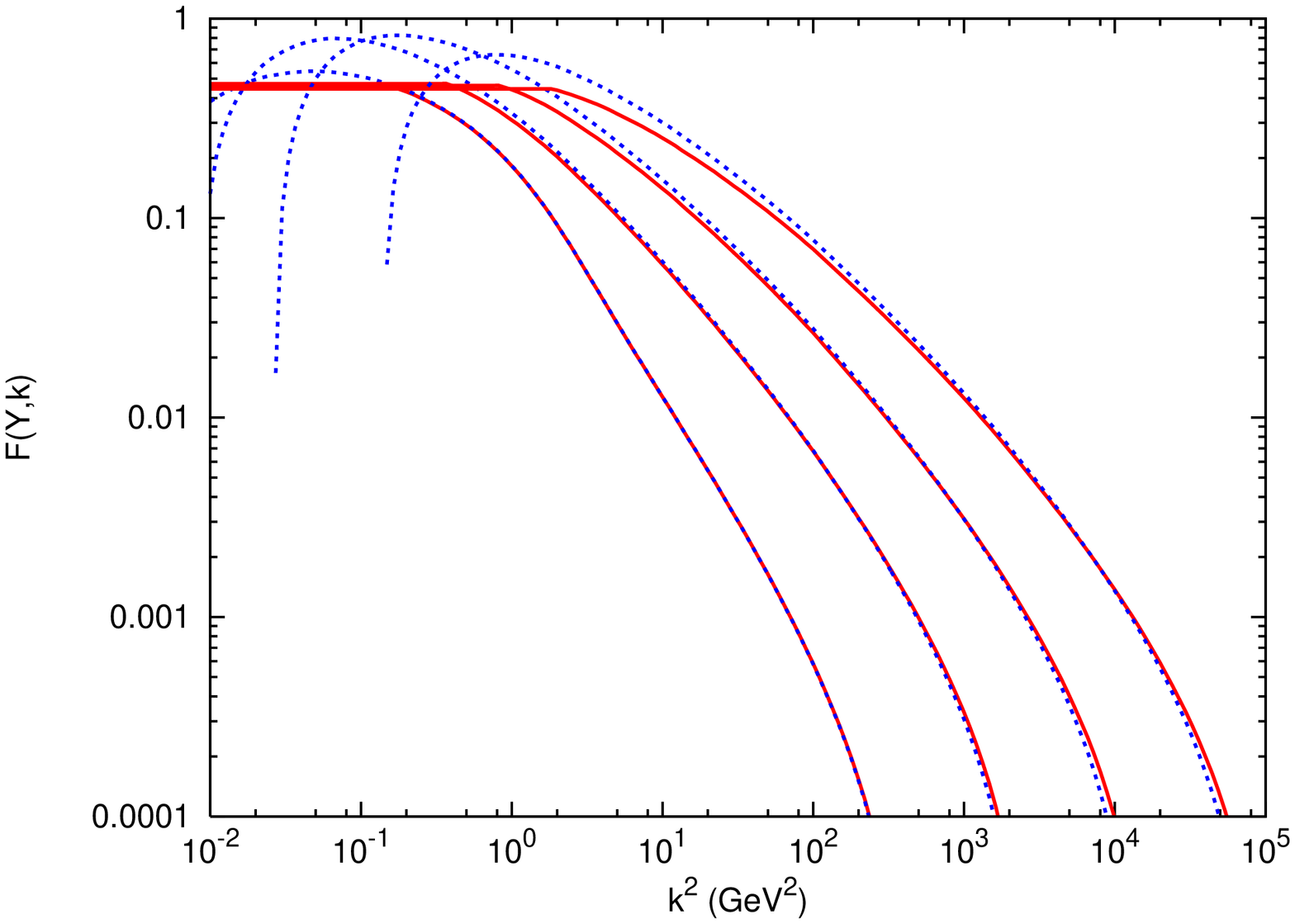}
\includegraphics[angle=270, scale=0.35]{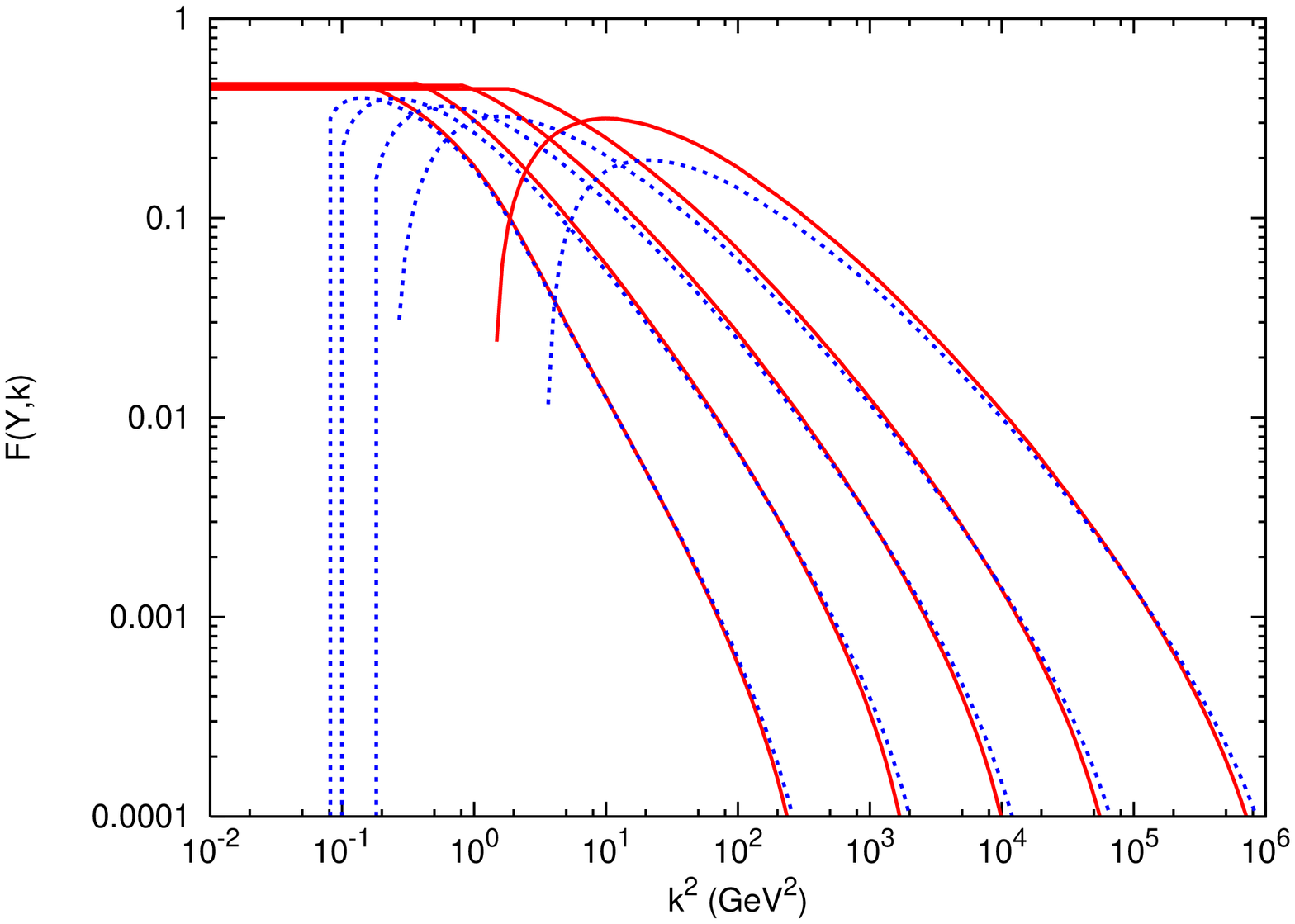}
\end{center} 
\caption{\label{fig:fixnlllinabs2} \emph{Left:} Comparison of the NLL BFKL solution with (solid red) and without (dotted blue)
the frozen saturation boundary \protect \eqref{frozenbound} for $Y=2,6,10,14$.  \emph{Right:} Comparison of the NLL solution 
with the frozen (solid red) saturation boundary \protect \eqref{frozenbound} versus the absorptive (dotted blue)
saturation boundary for $Y=2,6,10,14, 20$.
Simulations done for the fixed coupling $\abar = 0.2$.
}
\end{figure}

Note also that the linear and nonlinear solutions differ slightly even in the very large $k$ region. 
It seems that this is due to the fact that the nonlinear solution with the saturation boundary 
kills the contributions at small $k$ which via the NLL kernel contribute negatively 
to the high-$k$ part of the solution.  If this is indeed the case, we would for the second boundary condition \eqref{frozenbound} expect that the difference compared to the linear solution at larger $k$ is somewhat smaller 
since in that case the contributions below the critical point are not set to zero. 
We show the results for the boundary \eqref{frozenbound}  in the left plot of figure \ref{fig:fixnlllinabs2}
where we can see that this is indeed the case. 
In the right plot we instead compare the solutions obtained by the two boundary conditions. Note that  
when $Y$ is large enough, also the solution obtained using the boundary 
\eqref{frozenbound} turns negative at smaller $k$.  
Thus we see that the nonlinearities associated
with the mechanism of gluon saturation do not cure the unstable low-$k$ behavior completely, even 
when they by construction stabilize the low $k$ region. 
  
\subsubsection{The NLL saturation scale for a fixed coupling}

As is clear from above it is hard to define the saturation scale $Q_s$ beyond a
certain value of $Y$ because of the severe instabilities of the solution. The results for the saturation scale
presented for the fixed coupling case should therefore not be taken too seriously, but we here want to 
demonstrate the very large effects of the higher order corrections on the solution. We again mention
that the full nonlinear equation might be even more unstable so it is not clear whether the 
standard notion of the saturation momentum even makes sense for the chosen value of $\abar$.
\begin{figure}[t]
\begin{center}
\includegraphics[scale=1.3]{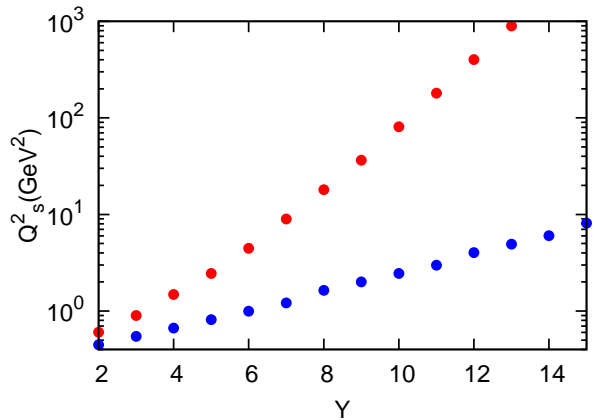}
\end{center} 
\caption{\label{fig:qsatfix} The saturation scale $Q_s^2(Y)$ extracted from the evolution using the 
boundary \protect \eqref{frozenbound}, for LL BFKL (red, upper circles) and NLL BFKL (blue, lower circles), with fixed coupling 
\protect $\abar = 0.2$.  }
\end{figure}  

The results are shown in figure \ref{fig:qsatfix} where we have extracted the saturation scale, $Q_s(Y)$, 
from the leading and next-to-leading order solutions using the boundary \eqref{frozenbound}. 
While the leading order evolution gives a very rapid, exponential, 
increase of $Q_s(Y)$ with $Y$, the next-to-leading order evolution leads to a strongly suppressed result. 
 Thus we can expect the NLL corrections 
to the BFKL kernel to have dramatic effects for the study of the nonlinear evolution as well, but the analysis cannot be 
complete until the important running coupling effects are taken into account. As we will see below, the running of the coupling has a rather large effect on the evolution and on the result presented above.

\subsection{Results with running coupling}

In the case of the running coupling we are faced with the question as to how exactly choose the 
scale of the coupling. For the scale dependence of $\alpha_s$ we shall use the one loop result 
resummed to all orders, that is 
\begin{equation}
\alpha_s(k) = \frac{1}{b \ln\bigl((k^2+\mu_{\rm IR}^2)/\Lambda^2\bigr)} \; , 
\end{equation}
where we have inserted an infrared regulator of the Landau pole. We shall by default set $\mu_{\rm IR}=0.7$ GeV.
We should immediately note that the linear evolution is sensitive to this parameter since a smaller $\mu_{\rm IR}$
implies an enhanced contribution from smaller momenta which generally speeds up the growth. To check the sensitivity to $\mu_{\rm IR}$ we have also run the simulations with 
$\mu_{\rm IR}=0.4$ GeV, which, as expected, speeds up the growth of the linear solutions, but we find that 
to a very good accuracy it does not affect the nonlinear solutions obtained from the saturation boundary
which are therefore rather robust with respect to the regulator. 
 
As can be seen from the NLL BFKL equation, the natural scale in the leading part of the kernel is given by 
the transverse momentum of the real gluon, $q$. The choice in the NLL part of the kernel is on the other hand rather arbitrary 
since any difference in the scale choice is formally of N$^2$LL and N$^3$LL order.  However, as we shall see,  
this formally higher order difference is extremely large and the solution therefore very sensitive to the exact 
choice. 

We have investigated different prescriptions for the running coupling. Let us write the BFKL kernel 
as in \eqref{bfklkernel} but this time extracting out the factors of $\abar$:
\begin{equation}
K(k,k')  =  \abar \, K_0(k,k')  + \abar^2 \, K_1(k,k').
\end{equation}
In studying the running coupling NLL evolution we have considered the following choices
($k_> \equiv \mathrm{max}(k,k')$ as before) 
\begin{align}
\text{A:} &\quad \abar(q^2) \, K_0(k,k')  + \abar^2(k_>^2) \, K_1(k,k') 
\label{scaleA}\\
\text{B:} &\quad \abar(k_>^2) \, K_0(k,k')  + \abar^2(k_>^2) \, K_1(k,k') \\
\text{C:} &\quad \abar(k^2) \, K_0(k,k')  + \abar^2(k^2) \, K_1(k,k') \\
\text{D:} &\quad \abar(q^2) \, K_0(k,k')  + \abar^2(q^2) \, K_1(k,k') 
\end{align}
where in choice D, the real momentum $q$ is used in all the real terms only (the virtual terms are diagonal 
in $k$). In all cases the kernel $K_1$ has been adjusted so that the expressions are the same at the NLL level.

We find that the choices A and B give rather well behaved solutions but that C and D lead to 
very unstable results which rapidly turn negative and oscillate over very large $k$ intervals. 
The differences between the choices A and B are not that large and so here we will 
only present results obtained from choice A. The fact that this choice gives a stable result is
consistent with the findings of \cite{Ciafaloni:2003rd} where the different scale choices of the running coupling 
were also investigated. We note that in this choice the NLL kernel does not contain any terms which depend on the beta function coefficient.

\begin{figure}[t]
\begin{center}
\includegraphics[width=0.47\textwidth]{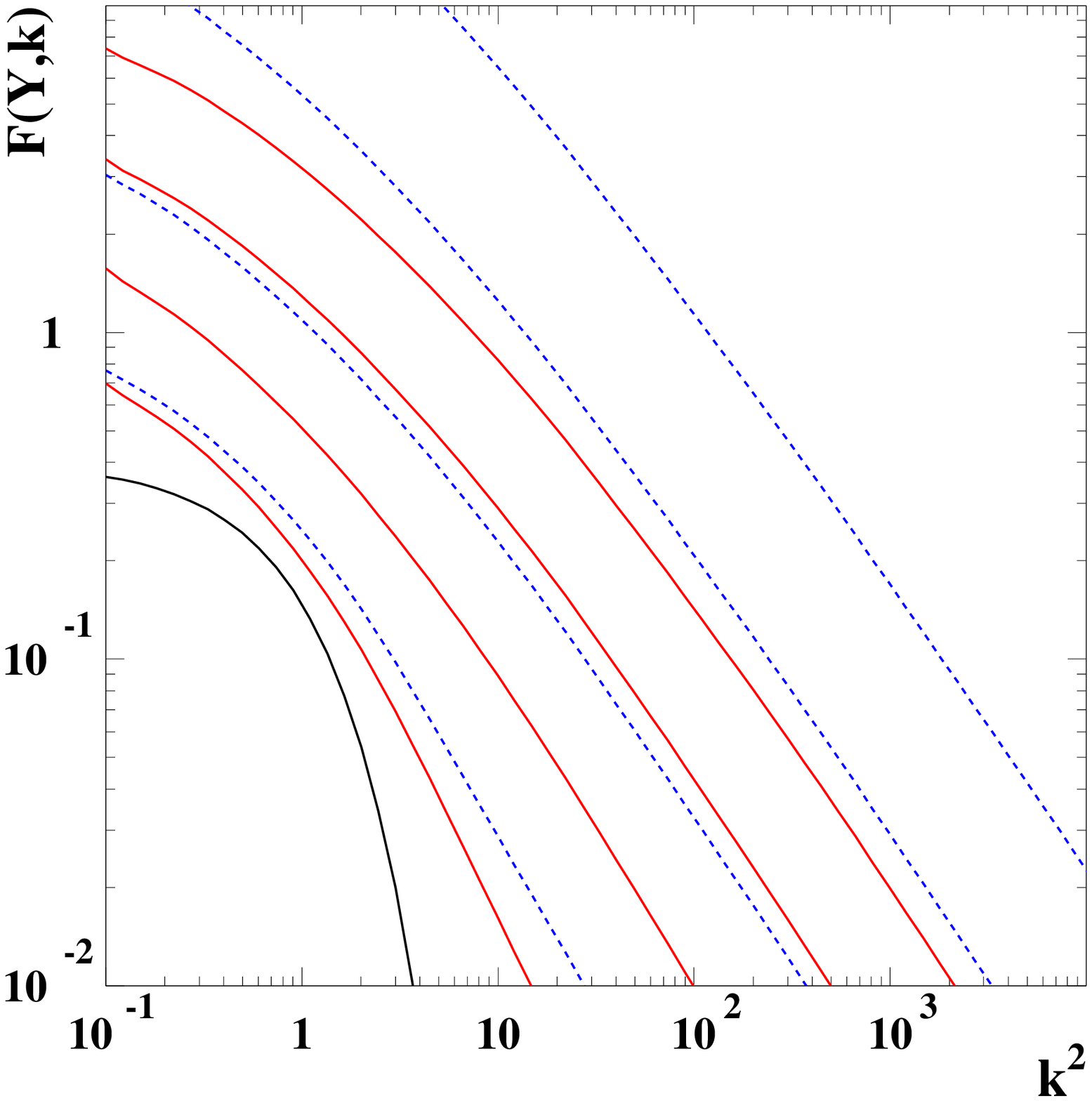}
\includegraphics[width=0.47\textwidth]{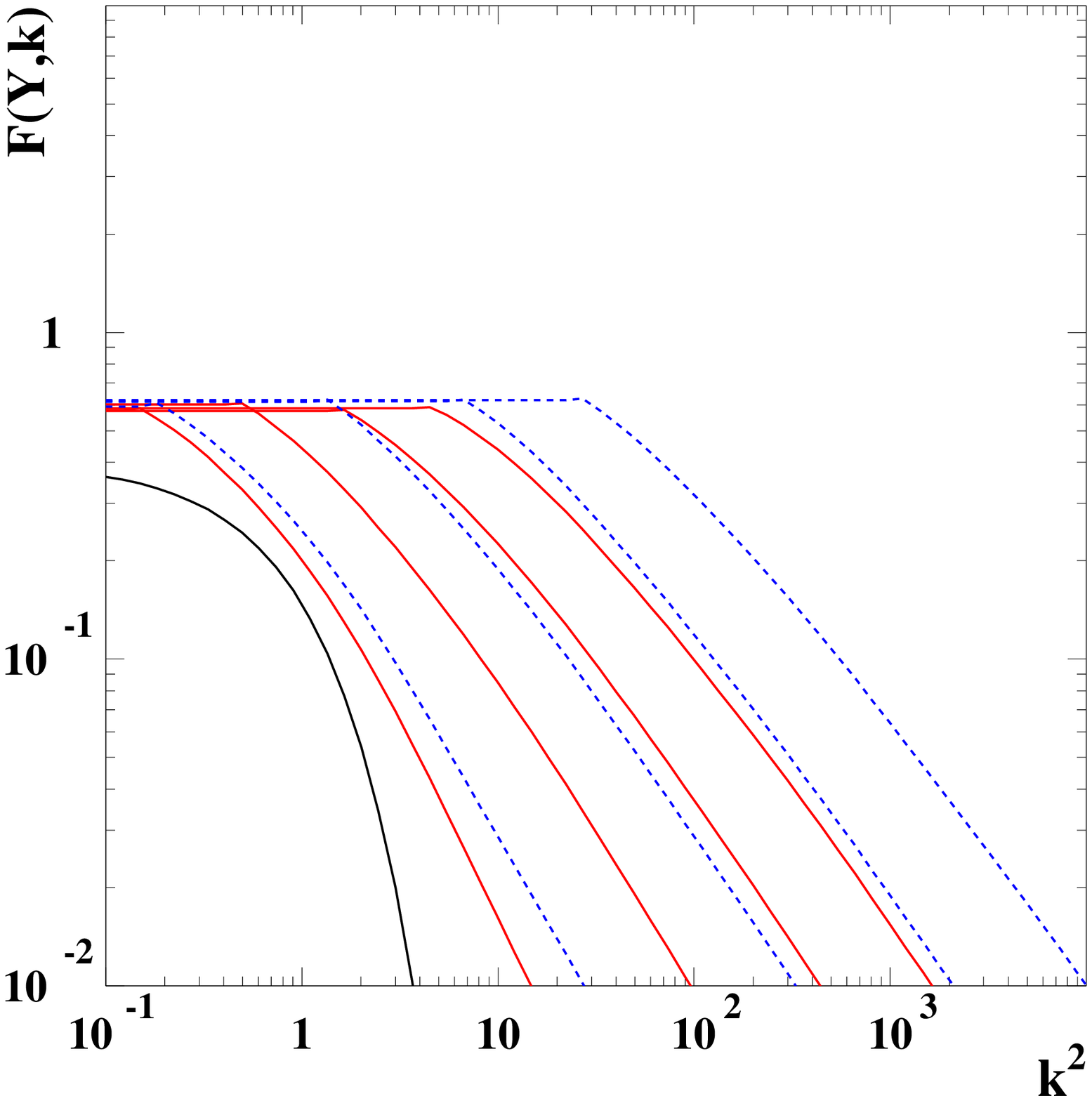}

\end{center} 
\caption{\label{fig:runalpha1} Full NLL BFKL solution (solid red) versus leading order running coupling solution (dotted blue)
with a running coupling using the scale choice A in \protect \eqref{scaleA} for $Y=2,6,10,14$.  Solid black curve is the initial condition at $Y=0$.
\emph{Left:} Linear evolution. \emph{Right:} Nonlinear evolution using the boundary \protect \eqref{frozenbound}.  
}
\end{figure} 

The results extracted using the scale choice A above are shown in figures \ref{fig:runalpha1} both for the linear and 
the nonlinear solutions.  For the linear case we again find a very large difference between the leading and the 
next-to-leading order solutions. The leading order solutions are here obtained using a running coupling which runs 
with the scale $q$ as clear from choice A in \eqref{scaleA}.  In this case the NLL solution seems better behaved 
than in the fixed coupling case studied in the previous section. It should, however, be kept in mind that while 
choice  A is stable, other choices like C and D lead to  unstable results. Thus the natural question arises whether  the NLL 
evolution including the running coupling  has any predictive power since it is extremely sensitive 
to corrections which are formally of higher order. 

\begin{figure}[t]
\begin{center}
\includegraphics[angle=0, width=0.47\textwidth]{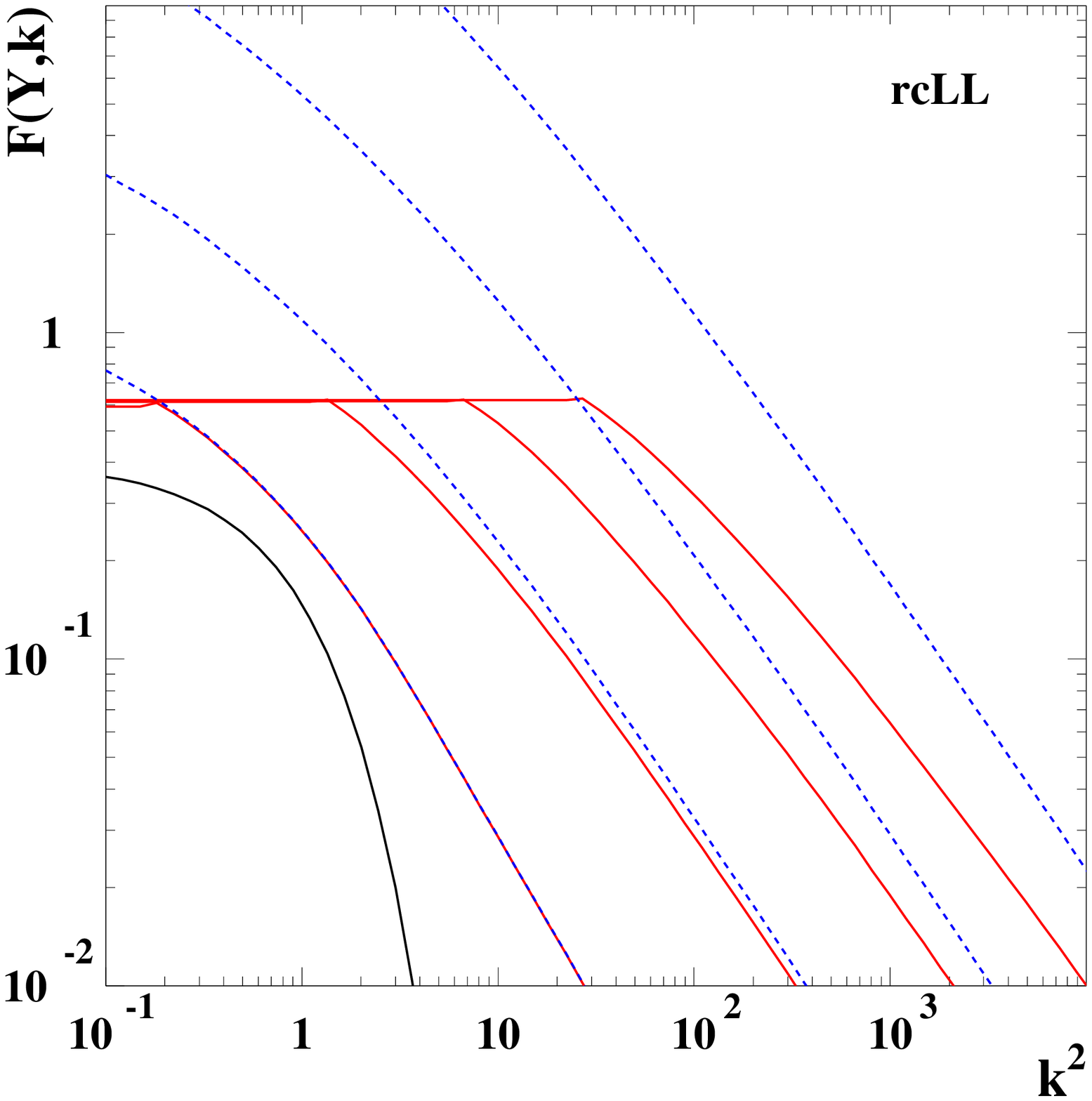}
\includegraphics[angle=0, width=0.47\textwidth]{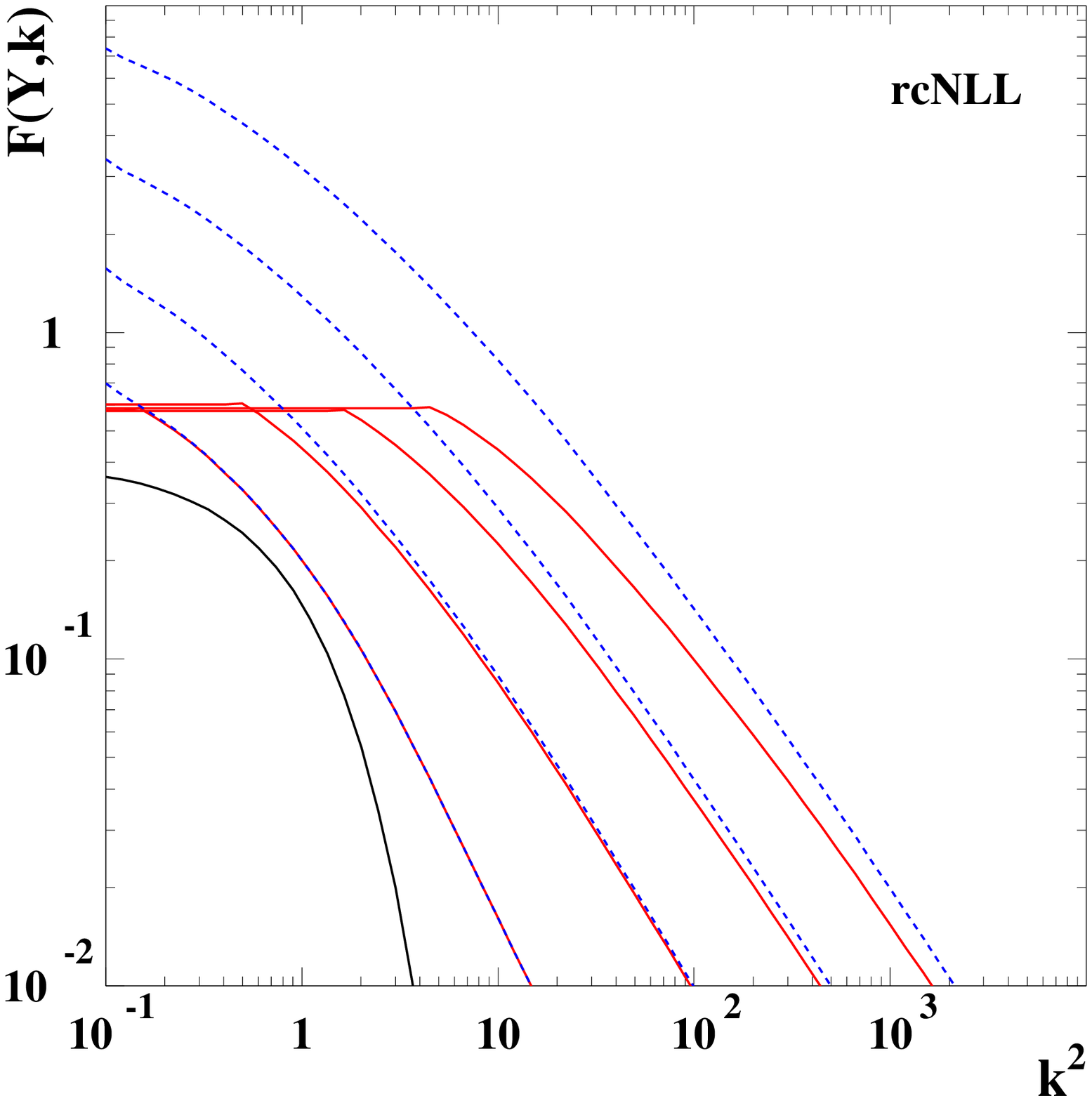}
\end{center} 
\caption{\label{fig:linvssat_lonlo} Linear BFKL solution (dotted blue) versus solution with the boundary (solid red)
with a running coupling using the scale choice A in \protect \eqref{scaleA} for $Y=2,6,10,14$.  
\emph{Left:} LL case. \emph{Right:} NLL case \protect \eqref{frozenbound}.  
}
\end{figure}

In the right plot in  figure  \ref{fig:runalpha1} we show the nonlinear solutions obtained after applying 
the frozen boundary condition \eqref{frozenbound}.  Apart from the solution at the lowest $Y$, we see that 
the running coupling leading order evolution (rcLL) has essentially the same slope as the full NLL solution 
but that it again grows rather more rapidly with $Y$. We also compare directly the linear and nonlinear solutions obtained 
from the boundary  in figure \ref{fig:linvssat_lonlo}.  In this plot, the differences between the solutions with saturation and the linear solutions are better visible. In the leading logarithmic case, application of the boundary affects the region far away from the boundary much more than in the next-to-leading scenario. This has a prominent effect on the speeds of the front evolution which in the next-to-leading case is much less affected by the saturation corrections.

The large sensitivity to the different scale choices in the running coupling is  not  unexpected. Given the large values of the running coupling the truncation of the small-$x$ perturbative expansion will lead to large uncertainties. These could be avoided by setting the scale choice via the BLM \cite{Brodsky:1982gc} scheme which reduces these uncertainties. Indeed, it has been demonstrated that the NLL small-x evolution can be stabilized in this scheme \cite{Brodsky:1998kn}. We note however that the resulting scale in the BLM scheme has a rather large numerical coefficient which reflects substantial differences between the $\overline{\rm MS}$ scheme and the BLM-MOM scheme adopted in \cite{Brodsky:1998kn}.
A related observation was done in \cite{Ivanov:2005gn}  where the principle of minimal sensitivity  
was applied to the NLL BFKL in the process of electroproduction of two vector mesons.
There it was found that the optimal choice of the renormalization scale is about $\sim 10 \, Q$
with $Q^2$ being the virtuality of the colliding photons \cite{Ivanov:2005gn,Ivanov:2006gt}.  One could explain this unnatural choice of scale by the fact that the higher order subleading corrections are effectively taken into account via this procedure. This was later confirmed by redoing the analysis \cite{Caporale:2007vs} using the collinearly improved, resummed kernel \cite{Vera:2005jt}, in which case the resulting optimal scale turns out to be $\sim 3 Q$,
which is much closer to the typical scales involved in the process. What all this shows is that the higher order
corrections are generally very large, and that the NLL evolution effectively stabilizes only by choosing a 
scale of the running coupling which is much larger than the natural choice dictated by the relevant physical 
scales in the process\footnote{It is indeed not strange that the NLL evolution would be well behaved in this case. 
Since the scale of the coupling is so large, and consequently its strength so small, the higher order corrections 
are automatically suppressed and presumably not important.}.

We also should add that the solutions in general are very sensitive to the lower cutoff on momentum $k$. In the simulations presented in this section we used the cutoff $k_{\rm min}^2=0.1 \; {\rm GeV}^2$. This sensitivity is of course due to the large value of the coupling  in this regime. It is worth noting, however, that the NLL solution does show some instability with respect to the variation of this cutoff.  Below $k_{\rm min}^2=0.05 \; {\rm GeV}^2$ we find that the low $k$ part of the solution turns 
negative and then oscillates very strongly leading to severe instabilities (also for the otherwise stable choices 
$A$ and $B$).  
While also the value of the LL solution (and resummed which we analyze later)  increases  with decreasing cutoff, 
in that case there is no instability of the solution as in the NLL case. We have therefore chosen the $k_{\rm min}$ here so that the NLL evolution gives stable results.
\subsubsection{The NLL saturation scale for a running coupling}

\begin{figure}[t]
\begin{center}
\includegraphics[scale=1.3]{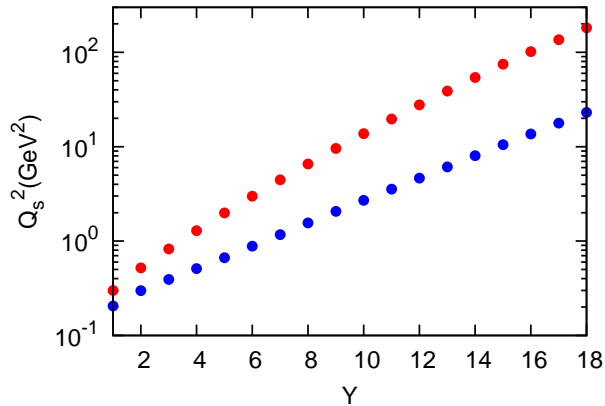}
\end{center}
\caption{\label{fig:qsatrun} The saturation scale $Q_s^2(Y)$ extracted using the boundary \protect \eqref{frozenbound} 
for the  running coupling 
solutions to the LL kernel (red, upper circles) and the full NLL kernel (blue, lower circles).  }
\end{figure}

We easily extract the saturation scale using the definition \eqref{qsatdef2} from the solution shown in figure 
\ref{fig:runalpha1}. The results are shown in figure \ref{fig:qsatrun} where the running coupling leading order results are
compared with the full NLL results. 
As we can see the differences between the leading order and the next-to-leading order kernels
are smaller than in the fixed coupling case, but we also note that they are again generally rather large.  At $Y=2$,
$Q_s$ differs by around a factor 1.5 between the running coupling leading result and the full NLL result, while
at $Y=18$ this difference has grown to a factor of almost 10.  Thus we clearly see that the next-to-leading 
order corrections beyond that of the running of the coupling are very important and cannot be neglected. 

Moreover, as mentioned above the scale choice A \eqref{scaleA} by which the results in  figure \ref{fig:qsatrun}
have been obtained is actually the more stable one giving a steady growth of $F$ with $Y$. While 
the  running coupling leading order solution does not give a markedly different $Q_s$ for other choices
of the scale of the coupling, the NLL solution changes dramatically, so that we cannot even sensibly 
extract a saturation scale. 

One might have already wondered what is the origin of these pathologies in the NLL evolution. To this end, it is instructive to look at the the NLL nonlinear equation in coordinate space, that is the NLL BK equation, for reasons that will be clear in what follows. The BK equation at leading order reads
 \begin{equation}\label{BK}
 \frac{\dif S_{\vec{x}\vec{y}}}{\dif Y} = 
 \int \dif^2 \vec{z}\, \mcal{M}^{(0)}_{\vec{x}\vec{y}\vec{z}}(S_{\vec{x}\vec{z}}S_{\vec{z}\vec{y}} - S_{\vec{x}\vec{y}}) \; ,
 \end{equation}
and has a straightforward interpretation. $S_{\vec{x}\vec{y}}$ is the $S$-matrix for the scattering of a color dipole $(\vec{x},\vec{y})$ of a generic hadronic target. Under an increment $\dif Y$ in rapidity, a soft gluon is emitted at the point $\vec{z}$ and we view it as a quark-antiquark pair at large $N_c$. Thus the parent projectile dipole can split into two, $(\vec{x},\vec{z})$ and $(\vec{z},\vec{y})$, which subsequently scatter off the target as suggested by the first term in the r.h.s., with the second corresponding to a self-energy correction. The probability for the splitting is of order $\mathcal{O}(\abar)$ and is given by the kernel $\mcal{M}^{(0)}_{\vec{x}\vec{y}\vec{z}}$ which is known and is positive for any value of its arguments. 

At NLL order the BK equation becomes (for example cf.~Eq.(104) in \cite{Balitsky:2010jf})
 \begin{equation}\label{NLOBK}
 \hspace*{-0.2cm}\frac{\dif S_{\vec{x}\vec{y}}}{\dif Y} = 
 \int \dif^2 \vec{z}\, \mcal{M}^{(1)}_{\vec{x}\vec{y}\vec{z}}(S_{\vec{x}\vec{z}}S_{\vec{z}\vec{y}} - S_{\vec{x}\vec{y}})
 +
 \int \dif^2 \vec{z}\, \dif^2 \vec{w}\, \widetilde{\mcal{M}}^{(1)}_{\vec{x}\vec{y}\vec{z}\vec{w}}(S_{\vec{x}\vec{z}}S_{\vec{z}\vec{w}}S_{\vec{w}\vec{y}} - \vec{w} \to \vec{z}),
 \end{equation}
where, for the sake of clarity, we have kept only the dominant terms at large $N_c$. Then the above again has a nice interpretation: the parent projectile dipole $(\vec{x},\vec{y})$ splits either into two dipoles $(\vec{x},\vec{z})$ and $(\vec{z},\vec{y})$ or into three $(\vec{x},\vec{z})$, $(\vec{z},\vec{w})$ and $(\vec{w},\vec{y})$, which then scatter off the hadronic target. The probabilities for these splittings to happen are given by the two kernels $\mcal{M}^{(1)}$ and $\widetilde{\mcal{M}}^{(1)}$, with $\mcal{M}^{(1)}$ containing the leading piece $\mcal{M}^{(0)}$ of order $\mathcal{O}(\abar)$ plus the NLL contribution of order $\mathcal{O}(\abar^2)$, while $\widetilde{\mcal{M}}^{(1)}$ is of order $\mathcal{O}(\abar^2)$. 

The problem is that a direct inspection of the two kernels reveals that they can both become negative and, moreover, large in magnitude. This will happen in the collinear and/or anticollinear limit, that is, when the emitted gluons at $\vec{z}$ and $\vec{w}$ are emitted very close to one of the parent color sources at $\vec{x}$ or $\vec{y}$, or very far from them. Since collinear splittings are related to the ultraviolet behavior of the evolution, we cannot expect saturation to cure this problem of large negative probabilities. We do not analyze this in more detail here, since in the next section we shall see how these pathologies arise in Mellin space.

Thus our main conclusion studying the NLL evolution in the presence of saturation effects is that 
the instabilities of the next-to-leading order BFKL evolution are not cured by the presence of the
nonlinear corrections. 
In order to obtain sensible results out of the evolution equations 
we must therefore consider an improvement that can cure the instabilities inherent in the formalism. 
As a specific model we here consider the resummation technique presented in \cite{Ciafaloni:2003rd}. 
We thus now turn to the renormalization group improved 
BFKL evolution which we shall subsequently study in the presence of saturation effects.

\section{Resummed BFKL with the boundary}
\label{resummedsection}

We begin this section by describing the resummation model presented in  \cite{Ciafaloni:2003rd}. As we have 
seen above, a resummation is needed in order to control the large higher order corrections in the small-$x$ evolution. 
After describing
the method we use, and the precise equation that it implies, we will go on to apply the saturation boundary 
and extract the resulting saturation scale. 


\subsection{Construction of resummed kernel}
Let us first briefly recall the principles of the resummation procedure originally presented in \cite{Ciafaloni:2003rd}, where more details can be found. Similar approaches to resummation have been developed in \cite{Altarelli:1999vw, Altarelli:2001ji, Altarelli:2003hk} with consistent results. The procedure consists of the resummation of the collinear singularities at NLL small-$x$ evolution, as well as in the incorporation of the running coupling.  
The eigenvalue (\ref{eq:nll}) contains double and triple collinear poles in Mellin space, which are numerically very large corrections.  Its approximate behavior near the  $\gamma\rightarrow 0$ and $\gamma\rightarrow 1$ poles is
\begin{equation}
\chi_1^{\rm coll}(\gamma) \sim \;  -\frac{1}{2\gamma^3}-\frac{1}{2(1-\gamma)^3}+\frac{A_1(0)}{\gamma^2}+\frac{A_1(0)-b}{(1-\gamma)^2} \; .
\label{eq:collinear}
\end{equation}
Here $A_1(0)=-11/12$ is coming from the 
 nonsingular part (in momentum fraction $z$) of the leading order DGLAP splitting function
\begin{equation}
\gamma_{gg}(\omega) = \asb \frac{1}{\omega} + \asb A_1(\omega) \; ,
\end{equation}
where Mellin transform has been defined as
\begin{equation}
\int_0^1 P_{gg}(z) z^{\omega} \dif z = \gamma_{gg}(\omega) \; .
\end{equation}
We shall also define 
\begin{equation}
\int_0^1 \tilde{P}_{gg}(z) z^{\omega} \dif z = \abar A_1(\omega) = \gamma_{gg}(\omega) - \frac{\abar}{\omega} \; .
\end{equation}

In \eqref{eq:collinear}, the double poles come from the DGLAP splitting function as mentioned above, whereas the triple collinear poles come from the choice of the energy scales.
The above collinear kernel (\ref{eq:collinear}) is negative and is responsible for  $90\%$ of the corrections of the NLL kernel in the case of the fixed coupling, that is when we set $b\rightarrow 0$. In order to smooth out the behavior near the collinear poles, the resummation procedure has to be taken into account. The two basic ingredients of this resummation are the subtraction of the above collinear poles and their replacement by the resummed expression which is due to the nonsingular part of DGLAP, and the kinematical constraint. That is, the resummed kernel reads
\begin{equation}
\chi_{\rm resum}(\gamma,\omega) = \chi_{0}(\gamma,\omega)+\chi_{\rm coll}(\gamma,\omega) + \bar{\alpha}_s \tilde{\chi}_1(\gamma) \; ,
\label{eq:resum1}
\end{equation}
with
\begin{equation}
\label{eq:resum2}
\begin{aligned}
\chi_0(\gamma,\omega) & = 2\psi(1)-\psi\bigl(\gamma+\omhalf\bigr)-\psi\bigl(1-\gamma+\omhalf\bigr)\; , \\
\chi_{\rm coll}(\gamma,\omega) & = \frac{\omega\,A_1(\omega)}{\gamma+\omhalf}+\frac{\omega\,A_1(\omega)}{1-\gamma+\omhalf}\; , \\
\tilde{\chi}_1(\gamma)  & = \chi_1(\gamma) +\frac{1}{2}\chi_0(\gamma) \frac{\pi^2}{\sin^2 (\pi \gamma)}-\chi_0(\gamma) \frac{A_1(0)}{\gamma(1-\gamma)} \; ,
\end{aligned}
\end{equation}
where the last line contains subtractions of the triple (second term) and double poles (third term) which numerically coincide with (\ref{eq:collinear}), modulo the $b$ dependent term. In contrast to the LL and NLL kernels, one of the important features of the resummed equation is that it satisfies energy conservation. 
Note that now the kernel eigenvalue is a function with both $\gamma$ and $\omega$ dependence. This is because the dependence on the coupling constant for some terms has been traded off for the dependence on $\omega$, due to the resummation. 

The above eigenvalue  is strictly correct for the fixed coupling case.  In the running coupling case, it is better to consider the momentum representation and perform the resummation of the coupling into the argument of the coupling in front of the leading order kernel.
Based on the resummed expression in Mellin space, the following proposal was made for the kernel in momentum space (we use again the notation $f$ to represent the action of the kernel as in \eqref{eq:llbfkl} and \eqref{eq:nllbfkl}):
\begin{multline}
 \int_x^1\frac{\dif z}{z}\int \dif \ktpsq \; \CKT(z;k,k') f\biggl(\frac{x}{z},k'\biggr) \\
 = \int_x^1\frac{\dif z}{z}\int \dif \ktpsq
 \Bigl[\bar{\alpha}_s(\qt^2) K_0^{\kc}(z;\kt,\kt') +
 \bar{\alpha}_s(k_{>}^2) K_{\ci}^{\kc}(z;k,k')+
 \bar{\alpha}^2_s({k}^2_{>}) \tilde{K}_1(k,k') \Bigr] f\biggl(\frac{x}{z},k'\biggr) \;.
 \label{eq:kernelxk}
\end{multline}
The different terms are as follows:
\begin{itemize}
\item LO BFKL with running coupling and consistency constraint
($\qt = \kt-\kt'$)
\begin{multline}
\label{eq:LOBFKLkc}
 \int_x^1\frac{\dif z}{z}\int \dif \ktpsq \; \Bigl[ \bar{\alpha}_s(q^2)
  K_0^{\kc}(z;\kt,\kt') \Bigr] f\biggl(\frac{x}{z},k'\biggr) \\
 = \int_x^1 \frac{\dif z}{z} \int\frac{\dif^2 \qt}{\pi q^2} \;
  \bar{\alpha}_s(q^2) \biggl[ f\biggl(\frac{x}{z},|\kt+\qt|\biggr)
 \Theta\biggl(\frac{k}{z}-k'\biggr)\Theta(k'-kz)-\Theta(k-q) f\biggl(\frac{x}{z},k\biggr) \biggr]\;.
\end{multline}
\item Nonsingular DGLAP terms with consistency constraint
\begin{multline}
 \int_x^1\frac{\dif z}{z}\int \dif \ktpsq \; \bar{\alpha}_s(k_{>}^2)
  K_{\ci}^{\kc}(z;k,k') f\biggl(\frac{x}{z},k'\biggr) \\
 = \int_x^1 { \dif z \over z} \int_{(kz)^2}^{k^2} \frac{\dif \ktpsq}{k^2} \;
  \bar{\alpha}_s(k^2) z\frac{k}{k'} \tilde{P}_{gg}\biggl(z\frac{k}{k'}\biggr)
  f\biggl(\frac{x}{z},k'\biggr)\\
 + \int_x^1 {\dif z \over z} \int_{k^2}^{(k/z)^2} \frac{\dif \ktpsq}{\ktpsq} \;
  \bar{\alpha}_s\bigl(\ktpsq\bigr) z\frac{k'}{k} \tilde{P}_{gg}\biggl(z\frac{k'}{k}\biggr)
  f\biggl(\frac{x}{z},k'\biggr) \;.  \label{eq:dglapterms}
\end{multline}
\item NLL part of the BFKL with subtractions included
\begin{multline}
 \int_x^1\frac{\dif z}{z}\int \dif \ktpsq \; \bar{\alpha}^2_s({k}^2_{>})
  \tilde{K}_1(k,k') f\biggl(\frac{x}{z},k'\biggr) 
 = \frac{1}{4}\int_x^1 \frac{\dif z}{z} \int \dif\ktpsq \; \bar{\alpha}^2_s({k}^2_{>}) \\
  \times\Biggl\{ 
 \biggl(\frac{67}{9} - \frac{\pi^2}{3}\biggr)\frac{1}{|\ktpsq-k^2|}
  \biggl[f\biggl(\frac{x}{z},\ktpsq\biggr) - \frac{2 k_{<}^2}{\ktpsq + k^2}f\biggl(\frac{x}{z},k^2\biggr)\biggr] \\
 +\Biggl[-\frac{1}{32}\Biggl(\frac{2}{\ktpsq} + \frac{2}{k^2} +
  \biggl(\frac{1}{\ktpsq} - \frac{1}{k^2}\biggr)
  \ln\biggl(\frac{k^2}{\ktpsq}\biggr)\Biggr) 
  + \frac{4 \Li\bigl(1-k_{<}^2/k_{>}^2\bigr)}{|\ktpsq - k^2|}\\
 {-4A_1(0)\operatorname{sgn}\bigl({k}^2-\ktpsq\bigr)
  \biggl(\frac{1}{k^2} \ln\frac{|\ktpsq-k^2|}{\ktpsq} - 
  \frac{1}{\ktpsq} \ln\frac{|\ktpsq-k^2|}{{k}^2}\biggr)} \\ 
 - \Biggl(3 + \biggl(\frac{3}{4} - \frac{(\ktpsq+k^2)^2}{32\ktpsq k^2}\biggr)\Biggr)
  \int_0^{\infty} \frac{\dif y}{k^2 + y^2 \ktpsq}
  \ln\biggl|\frac{1+y}{1-y}\biggr| \\ 
 + \frac{1}{\ktpsq + k^2} \biggl(\frac{\pi^2}{3} +
  4 \Li\biggl(\frac{k_{<}^2}{k_{>}^2}\biggr)\biggr)\Biggr]
  f\biggl(\frac{x}{z},k'\biggr) \Biggr\} 
 + \frac{6 \zeta(3)}{4}  \int_x^1 \frac{\dif z}{z} \;
  \bar{\alpha}^2_s(k^2)  f\biggl(\frac{x}{z},k\biggr) \;. 
\end{multline}
\end{itemize}
The fact that the shifts of the collinear poles are symmetric in (\ref{eq:resum2}) is reflected by the symmetric form of the kinematical constraint in (\ref{eq:LOBFKLkc}) and  (\ref{eq:dglapterms}). This in turn is related to the symmetric scale choice, 
$s_0=Q_AQ_B$. As mentioned above, in our calculation we actually use the asymmetric scale choice, 
$s_0=Q_A^2$.
In addition, in the kernel written above  one needs to perform additional subtractions. This is because
the kernel still contains some residual single poles.  They contribute to a residual 2-loop anomalous dimension which needs to be subtracted. For the purpose of this analysis we choose the scheme called scheme \textbf{B} in \cite{Ciafaloni:2003rd}.
It consists of  a modification which adds a term with  the shifted pole to the NLL kernel
with the $\om$-dependent coefficient
\begin{equation} \label{eq:subver3}
 \tilde{\chi}_1(\gamma) \to \tilde{\chi}_1^{\om}(\gamma) =
 \tilde{\chi}_1(\gamma) -
 \left( \frac1{\gamma} + \frac1{1-\gamma} \right) C(0) +
 \left( \frac1{\gamma+\omhalf} + \frac1{1+\omhalf-\gamma} \right)
 C(\om) [ 1+\om A_1(\om)] \;,
\end{equation}
where
\begin{equation}
\begin{aligned}
 C(\om) &= -\frac{A_1(\om)}{\om+1}
  +\frac{\psi(1+\om)-\psi(1)}{\om} \; , \\
 C(0) &= \frac{\pi^2 }{6}-A_1(0) \; .
\end{aligned}
\label{eq:sub1}
\end{equation}
 This scheme also satisfies the
energy-momentum sum rule for the extracted resummed  anomalous dimension $\gamma_{gg}$.

The change in the resummed kernel in $(x,k^2)$ space corresponding to scheme
\textbf{B} is obtained by taking the inverse Mellin transform of (\ref{eq:subver3}), and is
given by
\begin{multline}\label{eq:kerver3}
 \int_x^1\frac{\dif z}{z}\int \dif \ktpsq
  \Bigl\{ \bar{\alpha}_s(q^2)
  K_0^{\kc}(z;k,k')  + \bar{\alpha}_s(k_{>}^2)
  K_{\ci}^{\kc}(z;k,k') + \bar{\alpha}^2_s({k}^2_{>})\tilde{K}_1(k,k')  \Bigr\}
 f\biggl(\frac{x}{z},k'\biggr) - \\
 -\int_x^1 { \dif z \over z} \Biggl\{ C(0)\Biggl[ \int_0^{k^2}
 \frac{\dif \ktpsq}{k^2}
  \bar{\alpha}^2_s(k^2)  f\biggl(\frac{x}{z},k'\biggr)
+ \int_{k^2}^\infty  \frac{\dif \ktpsq}{\ktpsq}
 \bar{\alpha}^2_s(\ktpsq)  f\biggl(\frac{x}{z},k'\biggr) \Biggr] - \\
 - \Biggl[ \int_{(kz)^2}^{k^2}
\frac{\dif \ktpsq}{k^2}
 \bar{\alpha}^2_s(k^2)   z\frac{k}{k'} S_{2}\biggl(z\frac{k}{k'}\biggr) f\biggl(\frac{x}{z},k'\biggr)
+ \int_{k^2}^{(k/z)^2} \frac{\dif \ktpsq}{\ktpsq}
 \bar{\alpha}^2_s\bigl(\ktpsq\bigr)  z\frac{k'}{k} S_{2}\biggl(z\frac{k'}{k}\biggr) f\biggl(\frac{x}{z},k'\biggr) \Biggr] \Biggr\} \;,
\end{multline}
with the function $S_2(z)$ given by
\begin{multline}\label{d:S2}
 S_2(z) =\frac{1}{144 z} \Biggl\{ 132 + 24 {\pi}^2 + z
  \Bigl[-541 + 24 {\pi}^2 + 72 z (1 + 3 z)\Bigr]
  -  144 \ln\biggl(-1 + \frac{1}{z}\biggr) \ln\biggl(\frac{1}{z}\biggr) \\
 + 12 \Biggl( \ln(1 - z) \Bigl[-1 - 2 z \bigl(23 + z (-15 + 8 z)\bigr)
  - 12 (1 + z) \ln(1 - z)\Bigr] \\
  + 12 z \ln\biggl(-1 + \frac{1}{z}\biggr) \ln\biggl(\frac{1}{z}\biggr)
  + 2 z \bigl[1 + z (-21 + 5 z) - 6 \ln(1 - z)\bigr] \ln(z) -
  6 (-1 + 2 z) \ln^2(z)  \Biggr) \\
   + 144 (-1 + z) \Biggl[ \Li(z)+\frac{1}{2} \ln\biggl(\frac{1}{z}\biggr)
  \ln \biggl[ \frac{z}{(1-z)^2}
  \biggr] -\frac{{\pi}^2}{6} \Biggr] - 144 (1 + 2 z) \Li(1 - z) \Biggr\}.
\end{multline}

\subsection{Numerical results}

\begin{figure}[t]
\begin{center}
\includegraphics[width=0.6\textwidth]{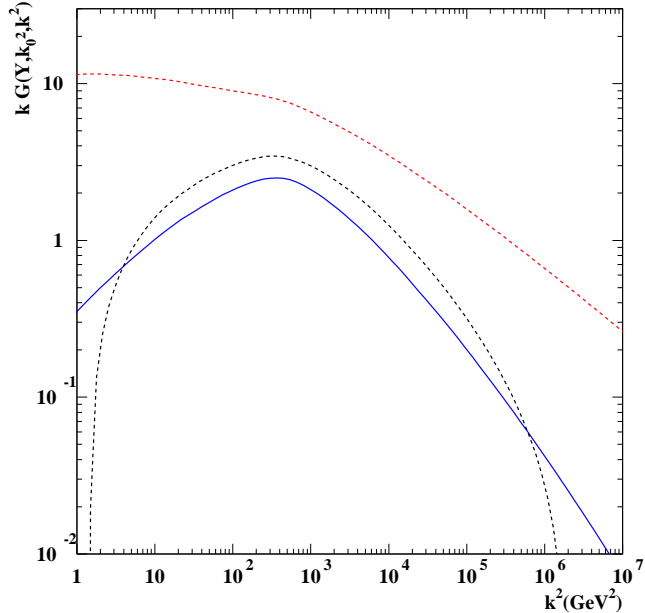}
\end{center}
\caption{The gluon Green's function solutions to the linear BFKL equation, obtained using a symmetric scale choice and a discrete delta initial condition. The red dashed line is the LL approximation, the black dashed is NLL, and the solid blue is the resummed prescription. In each case (including the LL case) the strong coupling constant is running as $\alpha_s(q)$ in front of the leading order term, and as $\alpha_s(\max(k,k'))$ in front of the subleading terms. The scale $k_0=20 \; {\rm GeV}$; all the solutions correspond to rapidity $Y=10$.}
\label{fig:nllvsresumlindel}
\end{figure}

In this section we present the numerical results for the evolution using the resummed kernel, 
also including the saturation boundary implemented via the conditions \eqref{absorpbound} and \eqref{frozenbound}
as before. Before going on to the study of the resummed kernel in the presence of the saturation boundary, however, 
let us first recall the comparison of the solutions for the linear LL, NLL and resummed kernels which was performed in \cite{Ciafaloni:2003rd}. In order to compare our different solutions to the gluon Green's function to those obtained in \cite{Ciafaloni:2003rd}, we use the symmetric scale and an initial distribution of the form of a discrete delta function in accordance with \eqref{bfklgreen} (rather than the initial condition \eqref{initialdistrb}). These results are shown in figure \ref{fig:nllvsresumlindel}. In all cases $\alpha_s(q)$ is taken as the choice for the running coupling in the leading term and $\alpha_s(\max(k,k'))$ for the subleading terms  (we recall 
that this corresponds to choice $A$ in \eqref{scaleA}).  The reduction of the NLL and resummed solutions with respect to the leading order solution is substantial in all regions of $k$. As we see, the resummed solution also has a better behavior in the large and small momentum limits as compared to the NLL solution. This  is due  to the absence of the double and triple collinear poles.

\begin{figure}[t]
\begin{center}
\includegraphics[width=0.47\textwidth]{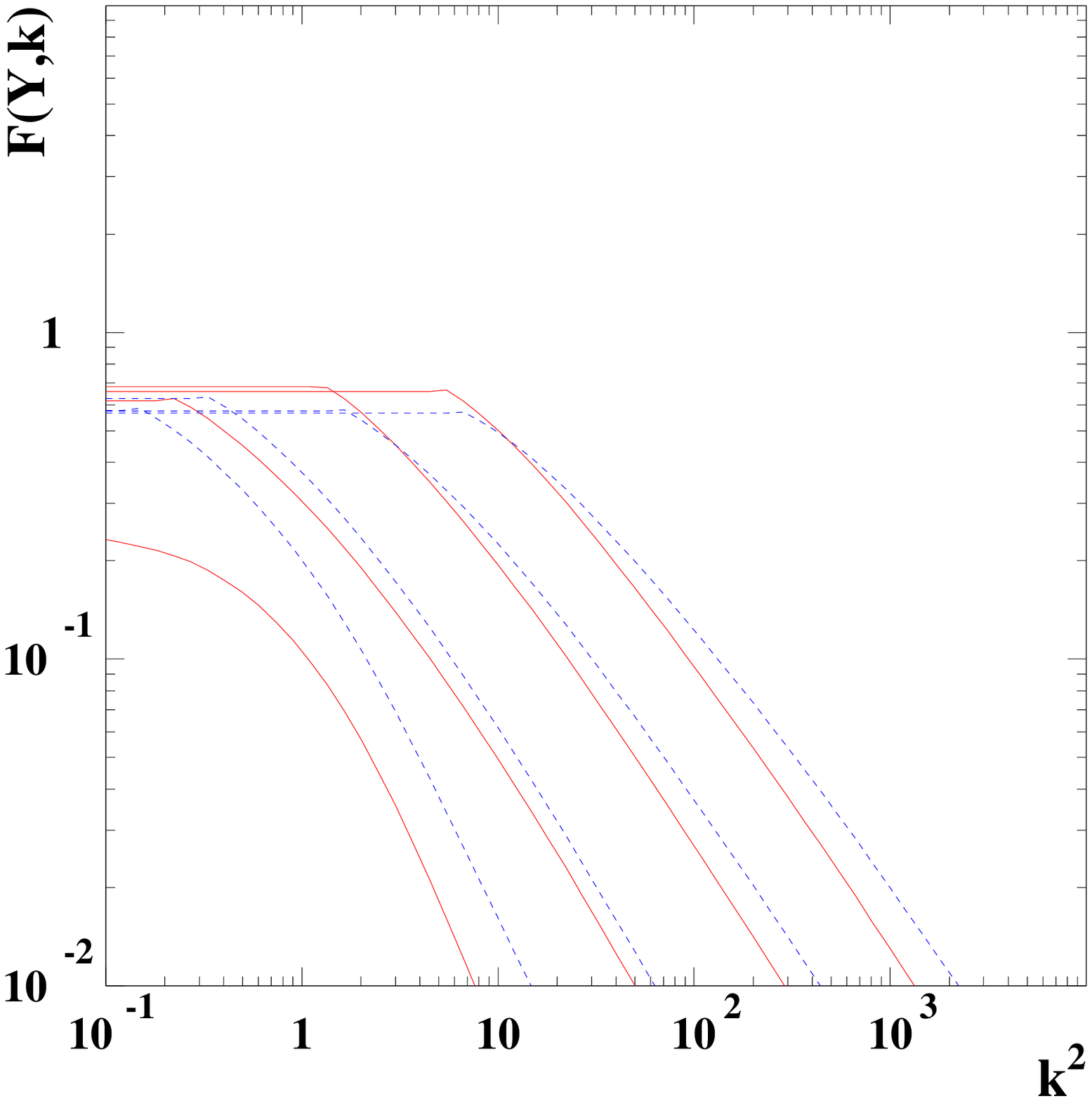}
\includegraphics[width=0.47\textwidth]{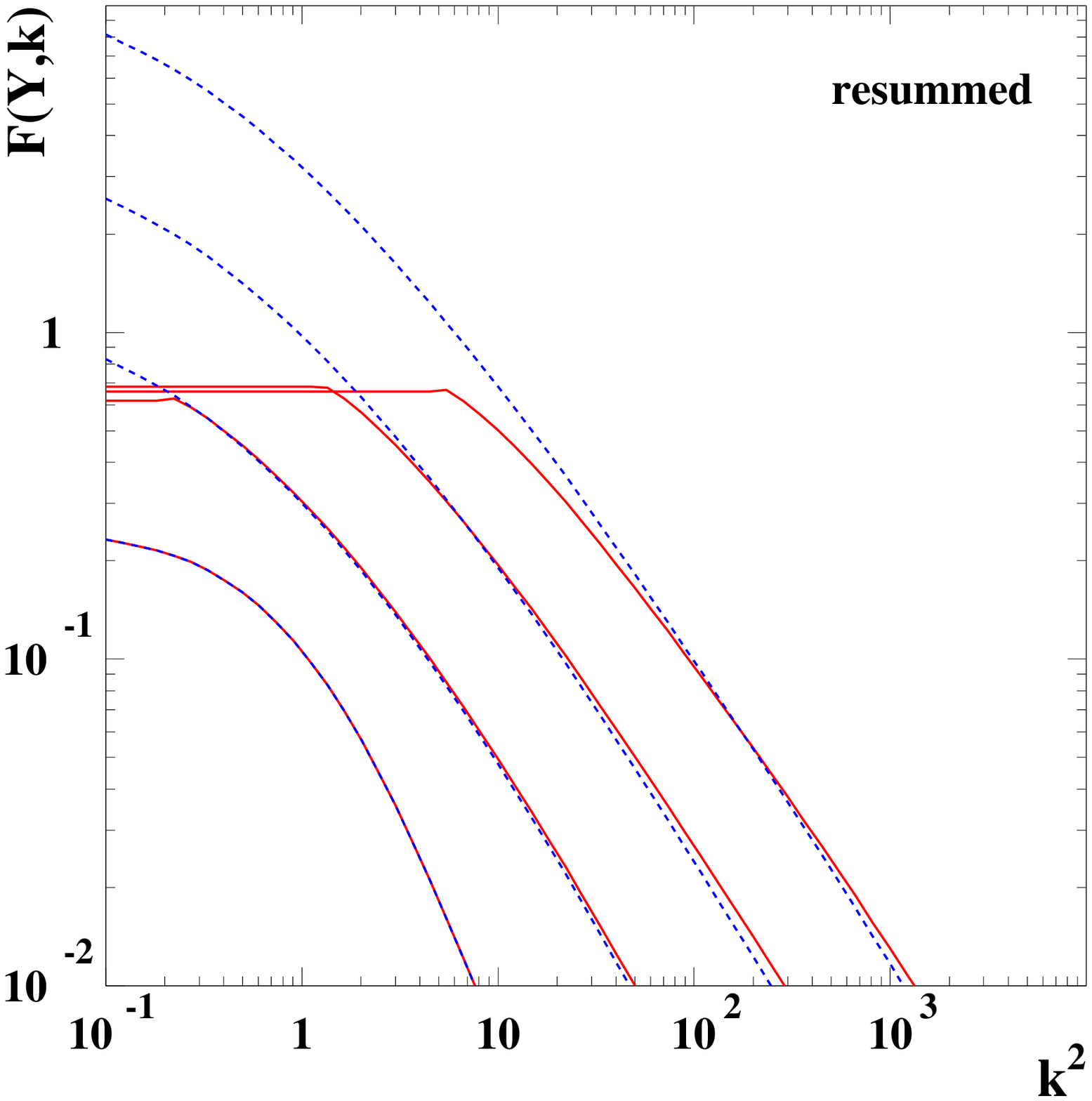}
\end{center}
\caption{\emph{Left:} The NLL (dotted blue) and resummed (solid red) BFKL evolutions with the saturation boundary \protect \eqref{frozenbound}. In each case the strong coupling constant is running as $\alpha_s(q)$ in front of the leading order term, and as $\alpha_s(\max(k,k'))$ in front of the subleading terms.  The four sets of curves correspond to $Y=2,6,10,14$. \emph{Right:} Linear resummed BFKL solution with (solid red) and without (dotted blue) the boundary \protect \eqref{frozenbound},
with a running coupling, for the same set of rapidities.}
\label{fig:nllvsresuminsat01}
\end{figure}

Next, we proceed to the analysis including the saturation boundary. For this,
as in section \ref{NLOresults},
we use the asymmetric scale choice and the initial condition \eqref{initialdistrb}.
The results are shown in figure \ref{fig:nllvsresuminsat01}. 
The resummed  solution grows initially rather slow which is to be expected as the resummed prescription contains a tower of terms which are subleading for high values of rapidity but which are nevertheless important for the phenomenologically relevant values of smaller $Y$. This delay in the growth of the resummed evolution has interesting and 
important consequences for the phenomenology of saturation as it clearly implies that the growth of the saturation 
scale will be delayed in $Y$.  

That the effects of saturation are thus suppressed in the resummed evolution for lower $Y$ is 
apparent in the right plot of figure \ref{fig:nllvsresuminsat01} where the nonlinear and linear evolutions
(i.e.~with and without the boundary, respectively)
are compared. Clearly, for the values of $Y$ shown in the figure, the progress of the front is not 
much affected by the inclusion of saturation.  This result can be compared to the earlier results 
presented for the LL and NLL evolutions in figure \ref{fig:linvssat_lonlo}.  It is then clear that saturation 
has a larger effect on the front in the NLL case than in the resummed case.  It is interesting that if we look 
at the linear solutions only (the dotted blue lines), then at the highest rapidity in the figure, $Y=14$, 
the resummed solution in  figure \ref{fig:nllvsresuminsat01} is actually even slightly larger than the NLL 
solution at the same rapidity in figure \ref{fig:linvssat_lonlo}. Despite this, however, the effect of saturation 
is manifestly smaller in the former. This is due to the fact that the evolution at low rapidities is significantly 
slowed down in the former due to the resummation, and this implies in turn that the saturation effects do not 
become important until later.  We also note that the 
resummed evolution is stable and there is therefore no complication or ambiguity in the extraction of 
the saturation scale. From the figures we also observe again the behavior at large $k$ which we noticed earlier in 
the case of the fixed coupling NLL evolution whereby the saturated solution actually lies slightly above the
linear one. The reason appears to be due to the collinear subtractions which still imply that the solution 
at very large $k$ can turn negative. Unlike the fixed coupling NLL evolution, however, the resummed 
evolution is stable, and its results are unambiguous. Moreover, the resummed prescription is not 
particularly sensitive to the choice of the running coupling, again unlike the NLL evolution which 
displays a high sensitivity.


\subsubsection{The  saturation scale from resummed approach}

It is straightforward to extract the saturation scale from the above results, but let us first look at the 
fixed coupling case where it is easier to make semi-analytic estimates of the result. We can then check the 
consistency of the implementation by comparing these estimates with the full numerical results.\\

\noindent\emph{Fixed coupling case}

It is possible to calculate analytically  the behavior of the saturation scale as a function of rapidity in the case of fixed coupling. This was originally done in \cite{Mueller:2002zm},
 and later also using the traveling wave method in \cite{Munier:2003sj}. The resummed evolution was also studied using the same methods in \cite{Triantafyllopoulos:2002nz}. 
 The full expression for fixed coupling is
\begin{equation}
  Q_s^2(Y) = Q_0^2 \exp\biggl(\frac{\bar\alpha_s\chi(\gamma_s,\bar\alpha_s)}{1 - \gamma_s}Y - \frac{3}{2(1 - \gamma_s)}\ln Y\biggr) \label{eq:satscale} \; .
\end{equation}
where the ``saturation anomalous dimension'' $\gamma_s$ is determined by the condition
\begin{equation}
\frac{\chi(\gamma_s,\bar\alpha_s)}{1-\gamma_s}=\chi'(\gamma_s,\bar\alpha_s).
\end{equation}
Here the prime on $\chi$ in the right hand side denotes the derivative with respect to $\gamma_s$.

Using this formula we can extract the rapidity dependence also for the resummed evolution. However, in this case the collinear resummation complicates somewhat the calculation because it introduces in the eigenfunction $\chi$ an additional dependence on the Mellin variable $\omega$, which is itself equal to $\bar\alpha_s\chi$. That is, we now have a  relationship,
 \begin{equation}
  \omega = \bar\alpha_s \chi_{\text{resum}}(\gamma,\bar\alpha_s,\omega) \label{eq:shifted} \;,
 \end{equation}
 where $\chi_{\mathrm{resum}}$ was given in \eqref{eq:resum1}. 
 This transcendental equation is not analytically solvable for $\omega$, but it can be numerically solved. We define a new function $\chi_\text{eff}(\gamma, \bar\alpha_s)$ such that for any values of $\gamma$ and $\bar\alpha_s$,
 \begin{equation}
  \omega = \bar\alpha_s \chi_\text{eff}(\gamma,\bar\alpha_s)
 \end{equation}
 gives the value of $\omega$ that satisfies equation \eqref{eq:shifted}. The solution $\chi_\text{eff}$ then fills the role of $\chi$ in equation \eqref{eq:satscale}, allowing us to calculate the rapidity dependence of the saturation scale.
As the absolute normalization of $Q_s$ is not under full control it is customary to calculate the 
logarithmic derivative
\begin{equation}
\lambda_s(Y)\equiv \frac{\dif \ln Q_s^2(Y)/Q_0^2}{\dif Y}.
\label{eq:lambdasat}
\end{equation}
It is well known that in the leading order BFKL evolution one has $\lambda_s=4.88\, \abar$,
and that $\gamma_s \approx 0.37$. 

Constructing the ``effective'' eigenfunction, $\chi_{\mathrm{eff}}$, numerically, we have calculated 
$\lambda_s$ for different values of $\abar$ for the resummed evolution. For rapidities up to 50 units 
we find that $\lambda_s$ is given by 0.308 for $\abar =0.1$ and 0.528 for $\abar=0.2$. If only the leading 
term in \eqref{eq:satscale} is kept then these numbers increase to 0.361 and 0.580 respectively. 
From the full numerical solution we instead find the numbers 0.322 and 0.558 respectively. Notice that 
these numbers are still rather below the leading order asymptotic result which indicates the importance 
of the non asymptotic corrections even up to relatively large $Y$. 
That the calculation slightly underestimates the slope obtained from simulation data seems reasonable because the highest order corrections that we are ignoring in the calculation are likely to be positive. In general, however, 
the calculation is rather consistent with the simulations.\\

\noindent\emph{The running coupling case and the full numerical solutions}
\begin{figure}[t]
\begin{center}
\includegraphics[width=0.6\textwidth]{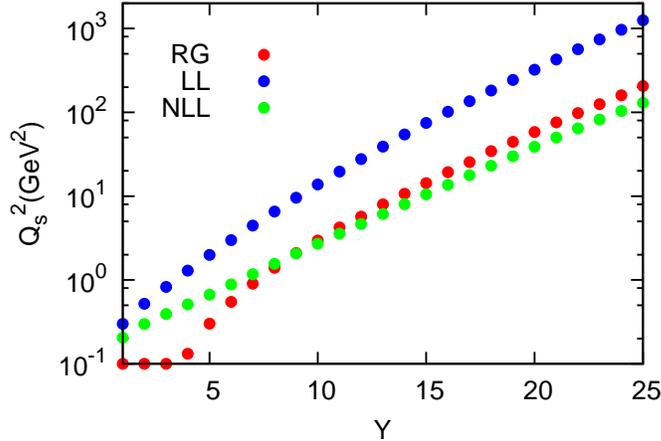}
\end{center}
\caption{The saturation scale obtained using the boundary \protect \eqref{frozenbound} as a function of rapidity for the LL (blue),  NLL (green) and resummed (red) evolutions. In each case the strong coupling constant is running as $\alpha_s(q)$ in front of the leading order term, and as $\alpha_s(\max(k,k'))$ in front of the subleading terms in the resummed and NLL cases. 
}
\label{fig:qsatresum}
\end{figure}

In figure \ref{fig:qsatresum} we present our main result, namely the saturation scale as a function of the rapidity for the LL, NLL  and resummed prescriptions. We here include only the results obtained using the boundary \eqref{frozenbound} for the resummed case. 
Let us mention that the different boundaries lead to different normalizations of the extracted saturation scales.
This difference in normalization  is due
to the evolution only for the lowest values of $Y$.  For the leading order solution for example, we find that the solutions 
obtained from different boundary prescriptions have the 
same $Y$ dependence for $Y \gtrsim 6$ (of course the exact numbers can depend on the initial condition). For the 
resummed evolution this number should be slightly higher due to the delayed evolution.  The overall differences
are not that large, however. 

We see from figure \ref{fig:qsatresum} that the saturation scale obtained from the resummed evolution
is suppressed at lower rapidities compared to the NLL result, though it should again be kept in mind that the 
NLL solution is generally unstable. 
In addition, also the absolute normalization of $Q_s$ cannot be taken too seriously since it depends
very much on the exact definition of $Q_s$.
What is clear, however, is that the resummed result is suppressed for the smallest 
rapidities due to the large-$x$ terms in the evolution. The ``plateau'' observed for the resummed saturation scale 
at lower rapidities in figure \ref{fig:qsatresum} can be attributed to the previously mentioned ``dip'' in the 
gluon splitting function. While the details of the results in figure \ref{fig:qsatresum} at the smallest $Y$
inevitably depend on  the exact initial condition and also the boundary condition (the type of the boundary 
and the numerical parameters chosen for given boundary), we have checked that the plateau of the 
saturation scale in the resummed
case also appears in case we change the boundary or when we consider a different initial condition (in particular
we have considered also an $x$-dependent initial condition which does not appear to wash out the plateau). 

\begin{figure}[t]
\begin{center}
\includegraphics[width=0.6\textwidth]{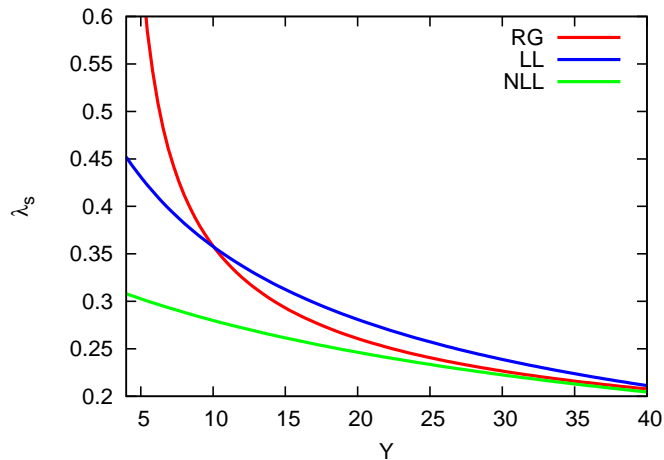}
\end{center}
\caption{The logarithmic derivative of the saturation momentum as a function of rapidity for  the 
LL (blue),  NLL (green), and resummed (red) evolutions obtained using the boundary condition 
\protect \eqref{frozenbound}.  In each case the strong coupling constant is running as $\alpha_s(q)$ in front of the leading order term, and as $\alpha_s(\max(k,k'))$ in front of the subleading terms in the resummed and NLL cases. 
}
\label{fig:qsatspeed3}
\end{figure}

We next show in figure \ref{fig:qsatspeed3} the logarithmic derivative of the saturation momentum, $\lambda_s$,
which was defined in  \eqref{eq:lambdasat}. 
We notice that the RG result is smaller than the LL one in the asymptotic region, and that it approaches it slowly from below in agreement with the results in \cite{Triantafyllopoulos:2002nz}. However, for values of $Y$ up to around 15, there is a discrepancy with the findings of \cite{Triantafyllopoulos:2002nz} which appears to be due to the presence of the non-asymptotic contributions in our numerical treatment. It is indeed clear from figure  \ref{fig:qsatspeed3} that the large-$x$ 
terms play a rather important role for the resummed result, as the red curve makes a sharp turn and exceeds the blue curve of the leading order result for smaller rapidities. Even though the ``speed'' of the saturation scale thus defined is relatively large for the resummed evolution for $Y \lesssim 10$, notice that the saturation scale itself is small as is evident from figure 
 \ref{fig:qsatresum}.

We thus see that the preasymptotic behavior of the resummed solution has important consequences for the 
study of saturation. How far exactly the plateau extends in $Y$ is hard to answer in our present application of the saturation 
boundary. To 
answer this question exactly, we would need to perform a serious phenomenological study where all the unknown 
parameters would be fitted to data. Regardless of the exact values of the parameters, however, it seems to be clear that 
resummation is needed for sensible results\footnote{It is, strictly speaking, not a problem if the gluon Green's function
goes negative at some $k$ since that does not automatically imply that the physical cross section goes negative. The Green's function
via formula \eqref{eq:sigmasym}, however, directly gives the energy dependence of the cross section which cannot have the unstable 
behavior observed att the NLL order. }, and that it implies that the growth of the saturation scale is delayed.

\section{Summary}
\label{sec:summary}

In this work we have studied small-$x$ evolution in the presence of saturation beyond the LL order. We have performed our analysis by self-consistently solving the linear BFKL evolution with proper boundary conditions; this is a justified procedure so long as we are interested in determining the energy dependence of the saturation momentum $Q_s$, and any observable under consideration for momenta above (but also close to) $Q_s$. We have found that the NLL corrections are large and that it is therefore not enough to keep only those next-to-leading terms related to the running of the coupling. Moreover, and as expected from a direct inspection of the NLL BK equation or from the well-known behavior of the NLL BFKL equation, the corrections can become negative in certain regimes of the transverse (momentum or coordinate) space and the NLL analysis is unstable. Thus we have used the renormalization group improved evolution kernel which resums all the dominant N$^n$LL contributions arising in the collinear (and anticollinear) limit, for all $n \geq 2$, and has a stable behavior. In accordance with previous expectations we do find that at high rapidities the running coupling results for the RG improved evolution converge to those of the LL evolution for the logarithmic derivative of the saturation momentum. However, for smaller values of the rapidity, that is for values up to $Y=10$ or 15, we find significant deviations from the asymptotic and the pure running coupling results. At least for the initial conditions studied, we find that the saturation momentum does not grow 
in the first few units of the rapidity interval, and  this is presumably the manifestation in the context of saturation of the well-known ``dip'' of the gluon-gluon splitting function. From that point on, $Q_s$ increases rather fast for the next few units until it becomes reliable to use the asymptotic expressions. At this stage, we do not know if our result is robust for all types of hadrons, e.g.~for a large nucleus, and if it is likely to change when one properly includes the relevant impact 
factors. However, it is an interesting phenomenon which, to our knowledge, has not been studied so far and clearly deserves further investigation.

\section*{Acknowledgments}
We would like to thank Guillaume Beuf and Edmond Iancu for valuable discussions and comments. 
This work is partially supported by U.S. D.O.E. grant number~DE-FG02-90-ER-40577, U.S. D.O.E. OJI  grant number DE-SC0002145 and   MNiSW  grant number
N202 249235.
A.M.S. is also supported by the Sloan Foundation.
\bibliographystyle{utcaps}
\bibliography{mybib}

\providecommand{\href}[2]{#2}\begingroup\raggedright\begin{thebibliography}{10}

\bibitem{Fadin:1975cb}
V.~S. Fadin, E.~A. Kuraev, and L.~N. Lipatov, ``{On the Pomeranchuk Singularity
  in Asymptotically Free Theories},'' {\em Phys. Lett.} {\bf B60} (1975)
50--52.

\bibitem{Balitsky:1978ic}
I.~I. Balitsky and L.~N. Lipatov, ``{The Pomeranchuk Singularity in Quantum
  Chromodynamics},'' {\em Sov. J. Nucl. Phys.} {\bf 28} (1978)
822--829.

\bibitem{Lipatov:1985uk}
L.~N. Lipatov, ``{The Bare Pomeron in Quantum Chromodynamics},'' {\em Sov.
  Phys. JETP} {\bf 63} (1986)
904--912.

\bibitem{Ciafaloni:1998gs}
M.~Ciafaloni and G.~Camici, ``{Energy scale(s) and next-to-leading BFKL
  equation},'' {\em Phys. Lett.} {\bf B430} (1998) 349--354,
\href{http://arXiv.org/abs/hep-ph/9803389}{{\tt hep-ph/9803389}}.

\bibitem{Camici:1997ij}
G.~Camici and M.~Ciafaloni, ``{Irreducible part of the next-to-leading BFKL
  kernel},'' {\em Phys. Lett.} {\bf B412} (1997) 396--406,
\href{http://arXiv.org/abs/hep-ph/9707390}{{\tt hep-ph/9707390}}.

\bibitem{Camici:1996st}
G.~Camici and M.~Ciafaloni, ``{Non-abelian q anti-q contributions to small-x
  anomalous dimensions},'' {\em Phys. Lett.} {\bf B386} (1996) 341--349,
\href{http://arXiv.org/abs/hep-ph/9606427}{{\tt hep-ph/9606427}}.

\bibitem{Fadin:1996nw}
V.~S. Fadin and L.~N. Lipatov, ``{Next-to-leading Corrections to the BFKL
  Equation From the Gluon and Quark Production},'' {\em Nucl. Phys.} {\bf B477}
  (1996) 767--808,
\href{http://arXiv.org/abs/hep-ph/9602287}{{\tt hep-ph/9602287}}.

\bibitem{Kotsky:1998ug}
M.~I. Kotsky, V.~S. Fadin, and L.~N. Lipatov, ``{Two-gluon contribution to the
  kernel of the Balitsky- Fadin-Kuraev-Lipatov equation},'' {\em Phys. Atom.
  Nucl.} {\bf 61} (1998)
641--656.

\bibitem{Fadin:1998py}
V.~S. Fadin and L.~N. Lipatov, ``{BFKL pomeron in the next-to-leading
  approximation},'' {\em Phys. Lett.} {\bf B429} (1998) 127--134,
\href{http://arXiv.org/abs/hep-ph/9802290}{{\tt hep-ph/9802290}}.

\bibitem{Salam:1999cn}
G.~P. Salam, ``{An Introduction to leading and next-to-leading BFKL},'' {\em
  Acta Phys.Polon.} {\bf B30} (1999) 3679--3705,
  \href{http://arXiv.org/abs/hep-ph/9910492}{{\tt hep-ph/9910492}}.

\bibitem{Salam:1998tj}
G.~P. Salam, ``{A resummation of large sub-leading corrections at small x},''
  {\em JHEP} {\bf 07} (1998) 019,
\href{http://arXiv.org/abs/hep-ph/9806482}{{\tt hep-ph/9806482}}.

\bibitem{Ciafaloni:1998iv}
M.~Ciafaloni and D.~Colferai, ``{The BFKL equation at next-to-leading level and
  beyond},'' {\em Phys.Lett.} {\bf B452} (1999) 372--378,
  \href{http://arXiv.org/abs/hep-ph/9812366}{{\tt hep-ph/9812366}}.

\bibitem{Ciafaloni:1999yw}
M.~Ciafaloni, D.~Colferai, and G.~P. Salam, ``{Renormalization group improved
  small-x equation},'' {\em Phys. Rev.} {\bf D60} (1999) 114036,
\href{http://arXiv.org/abs/hep-ph/9905566}{{\tt hep-ph/9905566}}.

\bibitem{Ciafaloni:2003rd}
M.~Ciafaloni, D.~Colferai, G.~P. Salam, and A.~M. Stasto, ``{Renormalisation
  group improved small-x Green's function},'' {\em Phys. Rev.} {\bf D68} (2003)
  114003,
\href{http://arXiv.org/abs/hep-ph/0307188}{{\tt hep-ph/0307188}}.

\bibitem{Altarelli:1999vw}
G.~Altarelli, R.~D. Ball, and S.~Forte, ``{Resummation of singlet parton
  evolution at small x},'' {\em Nucl.Phys.} {\bf B575} (2000) 313--329,
  \href{http://arXiv.org/abs/hep-ph/9911273}{{\tt hep-ph/9911273}}.

\bibitem{Altarelli:2001ji}
G.~Altarelli, R.~D. Ball, and S.~Forte, ``{Factorization and resummation of
  small x scaling violations with running coupling},'' {\em Nucl.Phys.} {\bf
  B621} (2002) 359--387, \href{http://arXiv.org/abs/hep-ph/0109178}{{\tt
  hep-ph/0109178}}.

\bibitem{Altarelli:2003hk}
G.~Altarelli, R.~D. Ball, and S.~Forte, ``{An Anomalous dimension for small x
  evolution},'' {\em Nucl.Phys.} {\bf B674} (2003) 459--483,
  \href{http://arXiv.org/abs/hep-ph/0306156}{{\tt hep-ph/0306156}}.

\bibitem{Vera:2005jt}
A.~Sabio~Vera, ``{An all-poles approximation to collinear resummations in the
  Regge limit of perturbative QCD},'' {\em Nucl. Phys.} {\bf B722} (2005)
  65--80,
\href{http://arXiv.org/abs/hep-ph/0505128}{{\tt hep-ph/0505128}}.

\bibitem{White:2006xv}
C.~White and R.~Thorne, ``{A Variable flavor number scheme for heavy quark
  production at small x},'' {\em Phys.Rev.} {\bf D74} (2006) 014002,
  \href{http://arXiv.org/abs/hep-ph/0603030}{{\tt hep-ph/0603030}}.

\bibitem{White:2006yh}
C.~White and R.~Thorne, ``{A Global Fit to Scattering Data with NLL BFKL
  Resummations},'' {\em Phys.Rev.} {\bf D75} (2007) 034005,
  \href{http://arXiv.org/abs/hep-ph/0611204}{{\tt hep-ph/0611204}}.

\bibitem{Iancu:2003xm}
E.~Iancu and R.~Venugopalan, ``{The Color glass condensate and high-energy
  scattering in QCD},'' \href{http://arXiv.org/abs/hep-ph/0303204}{{\tt
  hep-ph/0303204}}.

\bibitem{Balitsky:1995ub}
I.~Balitsky, ``{Operator expansion for high-energy scattering},'' {\em Nucl.
  Phys.} {\bf B463} (1996) 99--160,
\href{http://arXiv.org/abs/hep-ph/9509348}{{\tt hep-ph/9509348}}.

\bibitem{JalilianMarian:1997dw}
J.~Jalilian-Marian, A.~Kovner, and H.~Weigert, ``{The Wilson renormalization
  group for low x physics: Gluon evolution at finite parton density},'' {\em
  Phys. Rev.} {\bf D59} (1999) 014015,
\href{http://arXiv.org/abs/hep-ph/9709432}{{\tt hep-ph/9709432}}.

\bibitem{JalilianMarian:1997gr}
J.~Jalilian-Marian, A.~Kovner, A.~Leonidov, and H.~Weigert, ``{The Wilson
  renormalization group for low x physics: Towards the high density regime},''
  {\em Phys. Rev.} {\bf D59} (1999) 014014,
\href{http://arXiv.org/abs/hep-ph/9706377}{{\tt hep-ph/9706377}}.

\bibitem{Iancu:2000hn}
E.~Iancu, A.~Leonidov, and L.~D. McLerran, ``{Nonlinear gluon evolution in the
  color glass condensate. I},'' {\em Nucl. Phys.} {\bf A692} (2001) 583--645,
\href{http://arXiv.org/abs/hep-ph/0011241}{{\tt hep-ph/0011241}}.

\bibitem{Ferreiro:2001qy}
E.~Ferreiro, E.~Iancu, A.~Leonidov, and L.~McLerran, ``{Nonlinear gluon
  evolution in the color glass condensate. II},'' {\em Nucl. Phys.} {\bf A703}
  (2002) 489--538,
\href{http://arXiv.org/abs/hep-ph/0109115}{{\tt hep-ph/0109115}}.

\bibitem{Iancu:2001ad}
E.~Iancu, A.~Leonidov, and L.~D. McLerran, ``{The renormalization group
  equation for the color glass condensate},'' {\em Phys. Lett.} {\bf B510}
  (2001) 133--144,
\href{http://arXiv.org/abs/hep-ph/0102009}{{\tt hep-ph/0102009}}.

\bibitem{Kovchegov:1999yj}
Y.~V. Kovchegov, ``{Small-x F2 structure function of a nucleus including
  multiple pomeron exchanges},'' {\em Phys. Rev.} {\bf D60} (1999) 034008,
\href{http://arXiv.org/abs/hep-ph/9901281}{{\tt hep-ph/9901281}}.

\bibitem{Balitsky:2008zza}
I.~Balitsky and G.~A. Chirilli, ``{Next-to-leading order evolution of color
  dipoles},'' {\em Phys. Rev.} {\bf D77} (2008) 014019,
\href{http://arXiv.org/abs/0710.4330}{{\tt 0710.4330}}.

\bibitem{Balitsky:2009yp}
I.~Balitsky and G.~A. Chirilli, ``{High-energy amplitudes in N=4 SYM in the
  next-to-leading order},'' {\em Phys. Lett.} {\bf B687} (2010) 204--213,
\href{http://arXiv.org/abs/0911.5192}{{\tt 0911.5192}}.

\bibitem{Balitsky:2006wa}
I.~Balitsky, ``{Quark contribution to the small-$x$ evolution of color
  dipole},'' {\em Phys. Rev.} {\bf D75} (2007) 014001,
\href{http://arXiv.org/abs/hep-ph/0609105}{{\tt hep-ph/0609105}}.

\bibitem{Kovchegov:2006vj}
Y.~V. Kovchegov and H.~Weigert, ``{Triumvirate of running couplings in small-x
  evolution},'' {\em Nucl. Phys.} {\bf A784} (2007) 188--226,
\href{http://arXiv.org/abs/hep-ph/0609090}{{\tt hep-ph/0609090}}.

\bibitem{Mueller:2002zm}
A.~H. Mueller and D.~N. Triantafyllopoulos, ``{The energy dependence of the
  saturation momentum},'' {\em Nucl. Phys.} {\bf B640} (2002) 331--350,
\href{http://arXiv.org/abs/hep-ph/0205167}{{\tt hep-ph/0205167}}.

\bibitem{Triantafyllopoulos:2002nz}
D.~N. Triantafyllopoulos, ``{The energy dependence of the saturation momentum
  from RG improved BFKL evolution},'' {\em Nucl. Phys.} {\bf B648} (2003)
  293--316,
\href{http://arXiv.org/abs/hep-ph/0209121}{{\tt hep-ph/0209121}}.

\bibitem{Munier:2003vc}
S.~Munier and R.~B. Peschanski, ``{Geometric scaling as traveling waves},''
  {\em Phys.Rev.Lett.} {\bf 91} (2003) 232001,
  \href{http://arXiv.org/abs/hep-ph/0309177}{{\tt hep-ph/0309177}}.

\bibitem{Avsar:2009pv}
E.~Avsar and E.~Iancu, ``{BFKL and CCFM evolutions with saturation boundary},''
  {\em Phys. Lett.} {\bf B673} (2009) 24--29,
\href{http://arXiv.org/abs/0901.2873}{{\tt 0901.2873}}.

\bibitem{Avsar:2009pf}
E.~Avsar and E.~Iancu, ``{CCFM Evolution with Unitarity Corrections},'' {\em
  Nucl. Phys.} {\bf A829} (2009) 31--75,
\href{http://arXiv.org/abs/0906.2683}{{\tt 0906.2683}}.

\bibitem{Beuf:2010aw}
G.~Beuf, ``{Universal behavior of the gluon saturation scale at high energy
  including full NLL BFKL effects},''
  \href{http://arXiv.org/abs/1008.0498}{{\tt 1008.0498}}.

\bibitem{Balitsky:2008rc}
I.~Balitsky and G.~A. Chirilli, ``{Conformal kernel for NLO BFKL equation in N
  = 4 SYM},'' {\em Phys.Rev.} {\bf D79} (2009) 031502,
  \href{http://arXiv.org/abs/0812.3416}{{\tt 0812.3416}}.

\bibitem{Balitsky:2010jf}
I.~Balitsky, ``{High-energy amplitudes in the next-to-leading order},''
  \href{http://arXiv.org/abs/1004.0057}{{\tt 1004.0057}}.

\bibitem{Ciafaloni:2003kd}
M.~Ciafaloni, D.~Colferai, G.~P. Salam, and A.~M. Stasto, ``{The Gluon
  splitting function at moderately small x},'' {\em Phys.Lett.} {\bf B587}
  (2004) 87--94, \href{http://arXiv.org/abs/hep-ph/0311325}{{\tt
  hep-ph/0311325}}.

\bibitem{Camici:1997sh}
G.~Camici and M.~Ciafaloni, ``{Non-abelian q anti-q contributions to small-x
  anomalous dimensions},'' {\em Nucl. Phys. Proc. Suppl.} {\bf 54A} (1997)
155--159.

\bibitem{Camici:1996fr}
G.~Camici and M.~Ciafaloni, ``{Model (in)dependent features of the hard
  pomeron},'' {\em Phys. Lett.} {\bf B395} (1997) 118--122,
\href{http://arXiv.org/abs/hep-ph/9612235}{{\tt hep-ph/9612235}}.

\bibitem{Fadin:1996zv}
V.~S. Fadin, M.~I. Kotsky, and L.~N. Lipatov, ``{Gluon pair production in the
  quasi-multi-Regge kinematics},''
\href{http://arXiv.org/abs/hep-ph/9704267}{{\tt hep-ph/9704267}}.

\bibitem{Mueller:2004sea}
A.~H. Mueller and A.~I. Shoshi, ``{Small-x physics beyond the Kovchegov
  equation},'' {\em Nucl. Phys.} {\bf B692} (2004) 175--208,
\href{http://arXiv.org/abs/hep-ph/0402193}{{\tt hep-ph/0402193}}.

\bibitem{Iancu:2004iy}
E.~Iancu and D.~N. Triantafyllopoulos, ``{A Langevin equation for high energy
  evolution with pomeron loops},'' {\em Nucl. Phys.} {\bf A756} (2005)
  419--467,
\href{http://arXiv.org/abs/hep-ph/0411405}{{\tt hep-ph/0411405}}.

\bibitem{Triantafyllopoulos:2005cn}
D.~N. Triantafyllopoulos, ``{Pomeron loops in high energy QCD},'' {\em Acta
  Phys. Polon.} {\bf B36} (2005) 3593--3664,
\href{http://arXiv.org/abs/hep-ph/0511226}{{\tt hep-ph/0511226}}.

\bibitem{Golec-Biernat:1999qd}
K.~Golec-Biernat and M.~Wusthoff, ``Saturation in diffractive deep inelastic
  scattering,'' {\em Phys. Rev.} {\bf D60} (1999) 114023,
\href{http://arXiv.org/abs/hep-ph/9903358}{{\tt hep-ph/9903358}}.

\bibitem{Golec-Biernat:1998js}
K.~Golec-Biernat and M.~Wusthoff, ``Saturation effects in deep inelastic
  scattering at low Q**2 and its implications on diffraction,'' {\em Phys.
  Rev.} {\bf D59} (1999) 014017,
\href{http://arXiv.org/abs/hep-ph/9807513}{{\tt hep-ph/9807513}}.

\bibitem{Kutak:2004ym}
K.~Kutak and A.~Stasto, ``{Unintegrated gluon distribution from modified BK
  equation},'' {\em Eur.Phys.J.} {\bf C41} (2005) 343--351,
  \href{http://arXiv.org/abs/hep-ph/0408117}{{\tt hep-ph/0408117}}.

\bibitem{Iancu:2002xk}
E.~Iancu, A.~Leonidov, and L.~McLerran, ``{The colour glass condensate: An
  introduction},''
\href{http://arXiv.org/abs/hep-ph/0202270}{{\tt hep-ph/0202270}}.

\bibitem{Avsar:2010ia}
E.~Avsar and A.~M. Stasto, ``{Non-linear evolution in CCFM: The Interplay
  between coherence and saturation},'' {\em JHEP} {\bf 1006} (2010) 112,
  \href{http://arXiv.org/abs/arXiv:1005.5153}{{\tt arXiv:1005.5153}}.

\bibitem{Andersen:2003wy}
J.~R. Andersen and A.~Sabio~Vera, ``{The Gluon Green's function in the BFKL
  approach at next-to-leading logarithmic accuracy},'' {\em Nucl.Phys.} {\bf
  B679} (2004) 345--362, \href{http://arXiv.org/abs/hep-ph/0309331}{{\tt
  hep-ph/0309331}}.

\bibitem{Andersen:2003an}
J.~R. Andersen and A.~Sabio~Vera, ``{Solving the BFKL equation in the
  next-to-leading approximation},'' {\em Phys.Lett.} {\bf B567} (2003)
  116--124, \href{http://arXiv.org/abs/hep-ph/0305236}{{\tt hep-ph/0305236}}.

\bibitem{Brodsky:1982gc}
S.~J. Brodsky, G.~Lepage, and P.~B. Mackenzie, ``{On the Elimination of Scale
  Ambiguities in Perturbative Quantum Chromodynamics},'' {\em Phys.Rev.} {\bf
  D28} (1983) 228.

\bibitem{Brodsky:1998kn}
S.~J. Brodsky, V.~S. Fadin, V.~T. Kim, L.~N. Lipatov, and G.~B. Pivovarov,
  ``{The QCD pomeron with optimal renormalization},'' {\em JETP Lett.} {\bf 70}
  (1999) 155--160, \href{http://arXiv.org/abs/hep-ph/9901229}{{\tt
  hep-ph/9901229}}.

\bibitem{Ivanov:2005gn}
D.~Ivanov and A.~Papa, ``{Electroproduction of two light vector mesons in the
  next-to-leading approximation},'' {\em Nucl.Phys.} {\bf B732} (2006)
  183--199, \href{http://arXiv.org/abs/hep-ph/0508162}{{\tt hep-ph/0508162}}.

\bibitem{Ivanov:2006gt}
D.~Ivanov and A.~Papa, ``{Electroproduction of two light vector mesons in
  next-to-leading BFKL: Study of systematic effects},'' {\em Eur.Phys.J.} {\bf
  C49} (2007) 947--955, \href{http://arXiv.org/abs/hep-ph/0610042}{{\tt
  hep-ph/0610042}}.

\bibitem{Caporale:2007vs}
F.~Caporale, A.~Papa, and A.~Sabio~Vera, ``{Collinear improvement of the BFKL
  kernel in the electroproduction of two light vector mesons},'' {\em
  Eur.Phys.J.} {\bf C53} (2008) 525--532,
  \href{http://arXiv.org/abs/0707.4100}{{\tt 0707.4100}}.

\bibitem{Munier:2003sj}
S.~Munier and R.~B. Peschanski, ``{Traveling wave fronts and the transition to
  saturation},'' {\em Phys. Rev.} {\bf D69} (2004) 034008,
\href{http://arXiv.org/abs/hep-ph/0310357}{{\tt hep-ph/0310357}}.

\end{thebibliography}\endgroup

\end{document}